\documentclass[twocolumn]{aastex631}

\shorttitle{Keck/NIRSPEC studies of \ion{He}{1} in the atmospheres of WASP-52b and WASP-177b}
\shortauthors{Kirk et al.}

\begin{document}

\title{Keck/NIRSPEC studies of \ion{He}{1} in the atmospheres of two inflated hot gas giants orbiting K dwarfs: WASP-52b and WASP-177b}

\correspondingauthor{James Kirk}
\email{james.kirk@cfa.harvard.edu}

\author[0000-0002-4207-6615]{James Kirk}
\affil{Center for Astrophysics $\vert$ Harvard \& Smithsonian, 60 Garden Street, Cambridge, MA 02138, USA}

\author[0000-0002-2248-3838]{Leonardo A. Dos Santos}
\affil{Space Telescope Science Institute, 3700 San Martin Drive, Baltimore, MD 21218, USA}
\affil{Observatoire astronomique de l’Universit\'e de Gen\`eve, Chemin Pegasi 51, 1290 Versoix, Switzerland}

\author[0000-0003-3204-8183]{Mercedes L\'{o}pez-Morales}
\affil{Center for Astrophysics $\vert$ Harvard \& Smithsonian, 60 Garden Street, Cambridge, MA 02138, USA}

\author[0000-0003-4157-832X]{Munazza K.\ Alam}
\affil{Center for Astrophysics $\vert$ Harvard \& Smithsonian, 60 Garden Street, Cambridge, MA 02138, USA}
\affil{Carnegie Earth \& Planets Laboratory, 5241 Broad Branch Road NW, Washington, DC 20015, USA}

\author[0000-0002-9584-6476]{Antonija Oklop{\v{c}}i{\'c}}
\affil{Anton Pannekoek Institute for Astronomy, University of Amsterdam, 1090 GE Amsterdam, Netherlands}

\author[0000-0002-1417-8024]{Morgan MacLeod}
\affil{Center for Astrophysics $\vert$ Harvard \& Smithsonian, 60 Garden Street, Cambridge, MA 02138, USA}

\author[0000-0003-1957-6635]{Li Zeng}
\affil{Department of Earth and Planetary Sciences, Harvard University, Cambridge, MA 02138, USA}

\author[0000-0002-4891-3517]{George Zhou}
\affil{Centre for Astrophysics, University of Southern Queensland, West Street, Toowoomba, QLD 4350, Australia}



\begin{abstract}

We present the detection of neutral helium at 10833\,\AA\ in the atmosphere of WASP-52b and tentative evidence of helium in the atmosphere of the grazing WASP-177b, using high-resolution observations acquired with the NIRSPEC instrument on the Keck II telescope. We detect excess absorption by helium in WASP-52b's atmosphere of $3.44 \pm 0.31$\,\% ($11\sigma$), or equivalently $66 \pm 5$ atmospheric scale heights. This absorption is centered on the planet's rest frame ($\Delta v = 0.00 \pm 1.19$\,km\,s$^{-1}$). We model the planet's escape using a 1D Parker wind model and calculate its mass-loss rate to be $\sim 1.4 \times 10^{11}$\,g\,s$^{-1}$, or equivalently 0.5\,\% of its mass per Gyr. For WASP-177b, we see evidence for red-shifted ($\Delta v = 6.02 +/- 1.88$\,km\,s$^{-1}$) helium-like absorption of $1.28 \pm 0.29$\,\% (equal to $23 \pm 5$ atmospheric scale heights). However, due to residual systematics in the transmission spectrum of similar amplitude, we do not interpret this as significant evidence for He absorption in the planet's atmosphere. Using a 1D Parker wind model, we set a $3\sigma$ upper limit on WASP-177b's escape rate of $7.9 \times 10^{10}$\,g\,s$^{-1}$. Our results, taken together with recent literature detections, suggest the tentative relation between XUV irradiation and \ion{He}{1} absorption amplitude may be shallower than previously suggested. Our results highlight how metastable helium can advance our understanding of atmospheric loss and its role in shaping the exoplanet population.

\end{abstract}

\keywords{exoplanets -- exoplanet atmospheres -- transmission spectroscopy}


\section{Introduction} 
\label{sec:intro}

Atmospheric escape is thought to play a key role in carving the demographics of observed exoplanets, with both the lack of short-period Neptunes (the `Neptune Desert', e.g., \citealt{Mazeh2016}) and the bimodal radius distribution of sub-Neptunes (the `Radius Valley', \citealt{Fulton2017}) the likely end-results of atmospheric loss \citep[e.g.,][]{Kurokawa2014,Lopez2013,Owen2013,Owen2017,Owen2018,Allan2019,Hallatt2021}. However, it is important that we build the sample of exoplanets that are observed to be \textit{actively losing} their atmospheres so that we can measure mass-loss rates and understand how these depend on planetary and stellar parameters, while also improving our understanding of the physics of, and interaction between, planetary and stellar winds.

A new avenue to observe ongoing mass-loss was recently opened by the first detection of helium in an exoplanet's atmosphere \citep{Spake2018}. This triplet, which absorbs in the near-IR at 10833\,\AA, can be observed from the ground and thus offers significant advantages over UV observations of Lyman-$\alpha$, which was the primary method of observing atmospheric escape prior to 2018 \citep[e.g.,][]{Vidal-Madjar2003,Lecavelier2010,Ehrenreich2015,Bourrier2018}.  

Indeed, there have been approximately two dozen papers targeting exoplanetary helium since 2018 \citep[e.g.,][]{Allart2018,Nortmann2018,Kirk2020,Vissapragada2020,Zhang2021}. These studies have resulted in more than ten planets with bona fide detections of helium (see Appendix \ref{sec:lit_detections} for a full list). This sample of planets reveals that K-type stars are the most favorable for observations of helium since they have the necessary extreme-UV to mid-UV flux ratios to maintain a populated metastable helium state in an exoplanet's atmosphere \citep{Oklopcic2019}. Additionally, previous studies have reported tentative evidence that planets that receive more XUV irradiation show larger amplitude helium absorption \citep[e.g.,][]{Nortmann2018,Alonso-Floriano2019,dosSantos2020}.

In this paper, we present \ion{He}{1} observations of two inflated hot gas giants orbiting K-type stars: WASP-52b \citep{Hebrard2013} and WASP-177b \citep{Turner2019}.

\subsection{WASP-52b}

WASP-52b, discovered by \cite{Hebrard2013}, is an inflated hot Saturn ($\mathrm{R_P} = 1.253 \pm 0.027$\,$\mathrm{R_{Jup}}$, $\mathrm{M_P} = 0.434 \pm 0.024$\,$\mathrm{M_{Jup}}$, $\mathrm{T_{eq}} = 1315 \pm 26 $\,K, \citealt{Mancini2017}) orbiting a young and active K2 dwarf (age $=0.4^{+0.3}_{-0.2}$\,Gyr, $\log R'_{\mathrm{HK}} = -4.4 \pm 0.2$, \citealt{Hebrard2013}). 

Previous studies of the planet's atmosphere in transmission are broadly consistent with muted spectral features, likely due to clouds in the planet's atmosphere \citep{Kirk2016,Chen2017,Louden2017,Mancini2017,Alam2018,May2018}, however, water has been detected in the near-IR \citep{Bruno2018,Bruno2020}, and Na, K, and H$\alpha$ have been detected at high-resolution \citep{Chen2017,Chen2020}. Additionally, these previous studies have revealed in-transit light curve anomalies from the optical to the near-IR associated with the planet occulting stellar magnetic regions \citep{Kirk2016,Louden2017,Mancini2017,Bruno2018,May2018}, highlighting the active nature of the host. Furthermore, WASP-52b is a JWST GTO target for transit and eclipse observations (PIDs: 1201 and 1224).

In \cite{Kirk2020}, we identified WASP-52b as a promising target for studies of atmospheric escape via helium due to its low surface gravity, large atmospheric scale height, and K-type host. Recently, \cite{Vissapragada2020} presented a photometric transit observation of WASP-52b in a narrow filter (FWHM =  0.635\,nm) centered on the \ion{He}{1} triplet. In this filter, they measured the planet's transit depth to be $2.97 \pm 0.13$\,\%, which was $1.6\sigma$ deeper than the transit depth observed by \cite{Alam2018} between 898.5 and 1030.0\,nm. 

In this study, we present the first high-resolution observation and detection of \ion{He}{1} in WASP-52b's atmosphere, which extends over $66 \pm 5$ atmospheric scale heights ($H$).

\subsection{WASP-177b}

WASP-177b, discovered by \cite{Turner2019}, is another inflated hot gas giant ($\mathrm{R_P} = 1.58^{+0.66}_{-0.36}$\,$\mathrm{R_{Jup}}$, $\mathrm{M_P} = 0.508 \pm 0.038$\,$\mathrm{M_{Jup}}$, $\mathrm{T_{eq}} = 1142 \pm 32 $\,K) orbiting an old K2 dwarf (age $=9.7 \pm 3.9$\,Gyr). WASP-177b is in a grazing transit configuration with an impact parameter of $0.980^{+0.092}_{-0.060}$ \citep{Turner2019}. The WASP data reveals the stellar photometry to modulate with a period of $14.86 \pm 0.14$\,d and amplitude of $5 \pm 1$\,mmag, indicating stellar magnetic regions. There have been no further studies of this planet.

Similar to WASP-52b, we also identified WASP-177b as a promising target for helium studies in \cite{Kirk2020} due to the planet's low surface gravity and large scale height, and the K-type host star. 

In this study, we present the first atmospheric follow-up of WASP-177b. We see tentative hints of \ion{He}{1} absorption extending across $23 \pm 5$\,$H$, however, as we discuss in section \ref{sec:trans_spec}, we do not interpret this as significant evidence for \ion{He}{1} absorption.

The rest of the paper is structured as follows. In sections \ref{sec:obs} and \ref{sec:dr} we describe our observations and data reduction. In Section \ref{sec:da} we detail our data analysis and results, including our helium transmission spectra in Section \ref{sec:trans_spec}. In Section \ref{sec:1d_models} we present our 1D atmospheric escape modelling of WASP-52b and WASP-177b's \ion{He}{1} transmission spectra. We discuss our results in Section \ref{sec:discussion} and conclude in Section \ref{sec:conclusions}.

\section{Observations} \label{sec:obs}

We observed one transit of WASP-52b on 1 Aug 2020 and one transit of WASP-177b on 4 Oct 2020 with the NIRSPEC instrument on Keck II, as part of program N110 (PI: Kirk). This is the same instrument used in \cite{Kirk2020} for our $30\sigma$ detection of \ion{He}{1} in the atmosphere of WASP-107b, and which has also been used by \cite{Kasper2020}, \cite{Zhang2021,Zhang2022_TOI,Zhang2022_HD}, and \cite{Spake2021} for \ion{He}{1} searches.

We used the NIRSPEC-1 filter which covers the wavelength range of 0.947--1.121\,$\mu$m ($Y$-band) at a nominal spectral resolution of 25000. For our observations of WASP-52b, we opted not to use the `Thin' blocking filter which can introduce fringing in NIRSPEC \citep[e.g.,][]{Kasper2020}. However, for our observations of WASP-177b, we incorrectly used the Thin blocking filter which did indeed lead to fringing in WASP-177's spectra, which we corrected for in our data reduction (section \ref{sec:dr}). 

Before and after each transit, we obtained 11 darks (which include the bias), 11 lamp flats, and two arcs (each comprised of 10 co-added frames with Ne, Ar, Xe, and Kr arc lamps). For our observations of WASP-52b, we took an additional two arc spectra midway through our observations to check the stability of the wavelength solution. Having found these arc spectra to be consistent with those at the start and end of our observations, we did not repeat this step for WASP-177b. We therefore acquired a total of 22 darks and 22 flats for each night, with six arcs for WASP-52b and four for WASP-177b.

For WASP-52b ($J$ = 10.6), we obtained 26 spectra with an exposure time of 1000\,s for the first six spectra, where clouds were overhead, and 600\,s for the remaining spectra. We acquired these spectra over the course of 313\,min, and used an ABBA nod pattern to remove the sky background from our reduced spectra.

For WASP-177b ($J$ = 10.7), we obtained 56 spectra with an exposure time of 300\,s. We were able to use a shorter exposure time on this night due to better observing conditions. We acquired these spectra over the course of 310\,min, and again used an ABBA nod pattern.

\section{Data reduction} \label{sec:dr}

\subsection{Extracting the wavelength calibrated spectra}

To reduce our NIRSPEC data we used the NIRSPEC-specific REDSPEC software \citep{McLean2003,McLean2007}, which is written in IDL. This was the same software as we used in \cite{Kirk2020} to reduce our WASP-107b data.

In short, REDSPEC performs dark and bias subtraction, flat-fielding, bad pixel interpolation, and standard spectral extraction following the spatial rectification of tilted spectra on the detector. 

For the dark and flat field corrections we median-combined the 22 darks and 22 flats to create a master dark and master flat for each night. We restricted our analysis to spectral orders 70 (1.080--1.101\,$\mu$m) and 71 (1.065--1.086\,$\mu$m) since these are the orders that cover the helium triplet at 1.0833\,$\mu$m (wavelength in vacuum). For WASP-52b, we achieved an average SNR of 128 per pixel per exposure for order 70, and 134 for order 71. For WASP-177b these values were 94 and 97, respectively. For both nights, and for both orders, we used a fourth-order polynomial to correct for the tilted nature of the spectra on the detector.

The next step in REDSPEC is to perform the wavelength calibration, which we did using our arc lamp spectra. For WASP-52b, we found that a cubic polynomial was able to map the measured locations of the arc lines to the theoretical values to within 0.01 pixels. 

For WASP-177b, we were unable to get a satisfactory wavelength solution from the arc lamps. This was because we chose to keep the slit position angle fixed at 12$^{\circ}$ to avoid a nearby star falling within the slit. This had the effect of shifting the stellar spectra and arc lamps, making accurate wavelength calibration difficult from the arcs. For WASP-52b, we allowed the sky to rotate on the slit which gives the most precise radial velocities for NIRSPEC\footnote{\url{https://www2.keck.hawaii.edu/inst/nirspec/obs_procedures.html}}, and found the arc solution to be satisfactory. For WASP-177b, we instead used the OH emission lines in the science spectra for our wavelength calibration.

We then extracted the spectra in differenced AB nod pairs, to remove the sky background and OH emission lines, using an aperture of 15 pixels. For WASP-177's spectra, where we did use the `Thin' blocking filter (section \ref{sec:obs}), we additionally had to perform a fringing correction. We did this using REDSPEC's fringing correction tool, which involved manually identifying and removing peaks in the power spectra calculated for each stellar spectrum. The flux uncertainties were calculated by considering the photon noise, read noise, dark current, and sky background. 

Due to the intermittent clouds in the first six frames of WASP-52b's observations, the sky background brightness varied between the A and B nod positions. This meant that a straightforward A-B subtraction did not adequately remove the OH emission lines from these frames. We experimented by adding an extra step in our spectral extraction process, by fitting the sky background with polynomials in the cross-dispersion direction, but found that this led to greater noise in the stellar spectra due to the uncertainty in the sky polynomial fit. Ultimately we found that our sigma clipping step (section \ref{sec:pp}) was able to remove the residual OH emission from these six frames. We also note that the OH emission lines were well seperated from the \ion{He}{1} triplet for this observation (Figure \ref{fig:molecfit_model}).

\subsection{Post-processing with \texttt{iSpec} and \texttt{molecfit}}
\label{sec:pp}

Following the extraction of the wavelength-calibrated stellar spectra using REDSPEC, we then post-processed our data to continuum normalize our spectra, correct for residual wavelength shifts, and remove telluric (primarily H$_2$O) absorption.

To continuum normalize our data, we used \texttt{iSpec} \citep{ispec1,ispec2} to fit cubic splines to a portion of each order's wavelength range (10800--10950\,\AA\ for order 70 and 10700--10850\,\AA\ for order 71), and masked the \ion{He}{1} triplet from our continuum calculation. We focused on these portions to improve our continuum normalization in the vicinity of the \ion{He}{1} triplet (at 10833\,\AA) while leaving enough telluric absorption features to allow for accurate correction.

\begin{table*}
\caption{The system parameters of WASP-52b and WASP-177b used in the data reduction and analysis. These values are from \protect\cite{Hebrard2013}, \protect\cite{Mancini2017} and \cite{Alam2018} for WASP-52b, and \protect\cite{Turner2019} for WASP-177b. The systemic velocities are from Gaia DR2 \protect\citep{Gaia1,Gaia2}.}
\label{tab:system_params}
\centering

\begin{tabular}{lcccc} \hline
Parameter & Symbol & Unit & WASP-52b & WASP-177b \\ \hline
Time of mid-transit & $T_0$ & BJD & $2456862.79776 \pm 0.00016^a$ & $2457994.37140 \pm 0.00028 $ \\

Orbital period & $P$ & d & $1.74978119 \pm 0.00000052^a$ & $3.071722 ± 0.000001$ \\
Orbital inclination & $i$ & $\circ$ & $85.15 \pm 0.06^a$ & $84.14^{+0.66}_{-0.83}$ \\
Continuum transit depth & $(R_p/R_*)^2$ & & $0.02686 \pm 0.00016$ & $0.0185^{+0.0035}_{-0.0014}$ \\
Semi-major axis & $a$ & AU & $0.02643 \pm 0.00055^a$ & $0.03957 \pm 0.00058$ \\
Scaled semi-major axis & $a/R_*$ & & $7.23 \pm 0.03^a$ & $9.61^{+0.42}_{-0.53}$ \\
Stellar mass & $M_*$ & \,$\mathrm{M}_{\odot}$ & $0.804 \pm 0.050^a$ & $0.876 \pm 0.038$ \\
Planet mass & $M_p$ & \,$\mathrm{M_J}$ & $0.434 \pm 0.024^a$ & $0.508 \pm 0.038$ \\
Planet radius & $R_p$ & \,$\mathrm{R_J}$ & $1.253 \pm 0.027^a$ & $1.58^{+0.66}_{-0.36}$ \\
Planet surface gravity & $\log g_p$ & cgs & $2.84 \pm 0.02^a$ & $2.67^{+0.22}_{-0.31}$ \\
Planet equilibrium temperature & $T_\mathrm{eq}$ & K & $1315 \pm 26^a$ & $1142 \pm 32$ \\
Semi-amplitude & $K_*$ &  m\,s$^{-1}$ & $84.3 \pm 3.0^b$ & $77.3\pm5.2$ \\
Systemic velocity & $\gamma$ & \,km\,s$^{-1}$ & $0.48 \pm 0.33^d$ & $-6.41\pm1.18^d$ \\ 
Stellar effective temperature & $T_\mathrm{eff}$ & K & $5000 \pm 100^b$ & $5017 \pm 70$ \\
Stellar metallicity & [Fe/H] & dex & $0.03 \pm 0.12^b$ & $0.25 \pm 0.04$ \\
Stellar surface gravity & $\log g$ & cgs & $4.553 \pm 0.010^a$ & $4.486 \pm 0.049$ \\
\hline
\multicolumn{5}{l}{$^a$\cite{Mancini2017}. $^b$\cite{Hebrard2013}. $^c$\cite{Alam2018}. $^d$Gaia DR2 \protect\citep{Gaia1,Gaia2}.} \\
\end{tabular}
\end{table*}

To remove the telluric absorption, we used \texttt{molecfit} \citep{molecfit1,molecfit2}, that has been used in a number of high-resolution ground-based transmission spectroscopy studies \citep[e.g.,][]{Allart2017,Allart2019,Nortmann2018,Kirk2020}. \texttt{molecfit} uses Global Data Assimilation System\footnote{\url{https://www.ncdc.noaa.gov/data-access/model-data/model-datasets/global-data-assimilation-system-gdas}} (GDAS) profiles which contain weather information for user-specified observatory coordinates, airmasses, and times. It then models the telluric absorption lines in the observed spectra using this information.

For WASP-52, order 70, we used six telluric absorption lines, free of significant stellar absorption, to constrain the \texttt{molecfit} model. For WASP-177, order 70, we used four telluric absorption lines to constrain the \texttt{molecfit} model.

\begin{figure}
    \centering
    \includegraphics[scale=0.375]{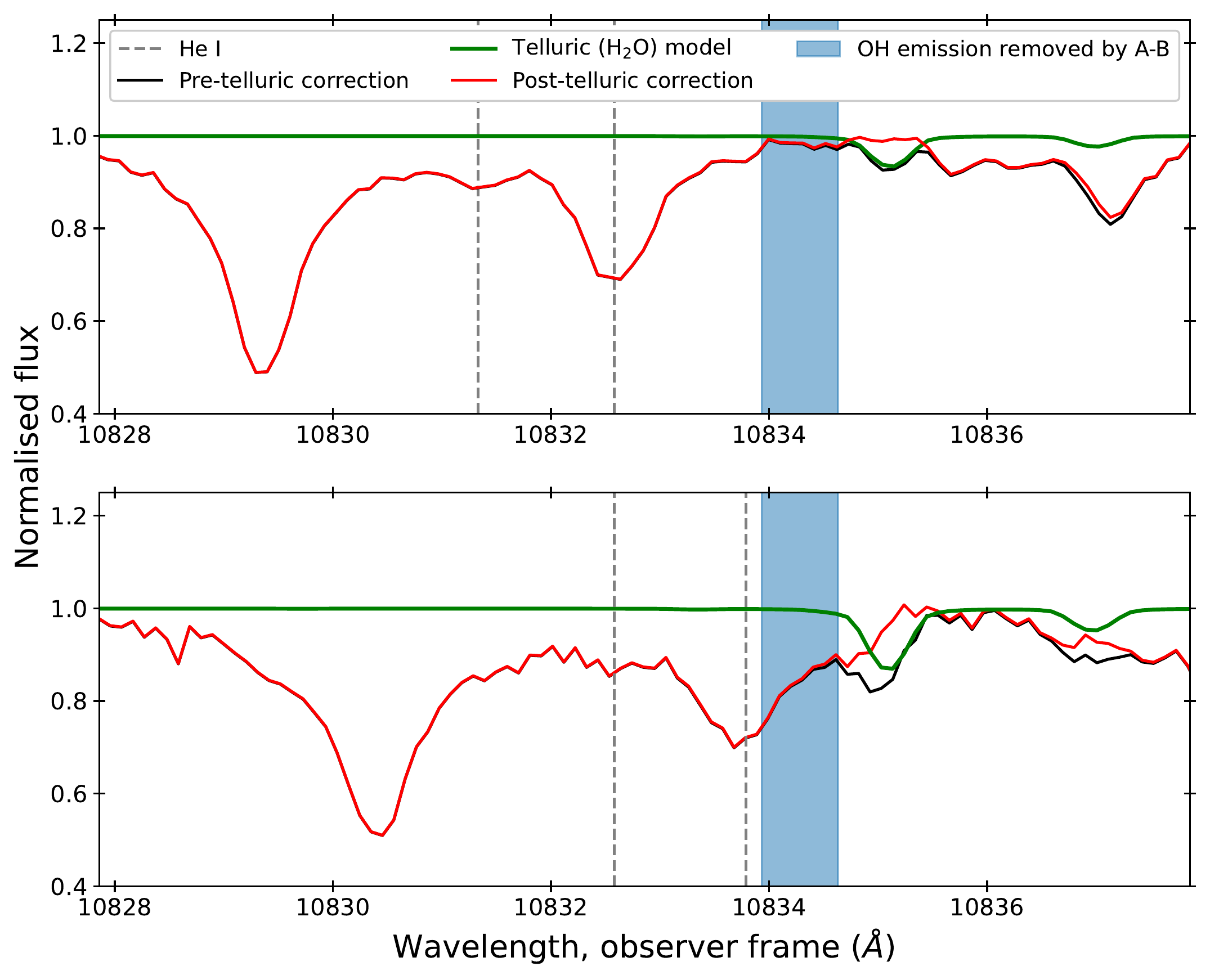}
    \caption{An example stellar spectrum of WASP-52 (top panel) and WASP-177 (bottom panel) before and after the telluric correction using \texttt{molecfit} \protect\citep{molecfit1,molecfit2}. The black spectra are the pre-corrected data, with the post-corrected data in red. The telluric models are shown in green with the shaded blue region indicating the wavelengths of strong OH emission lines that are removed by our differenced AB nod pairs. The helium triplet is denoted by the vertical dashed gray lines.}
    \label{fig:molecfit_model}
\end{figure}

We chose to fit only for the atmospheric H$_2$O content, with CH$_4$ and O$_2$ fixed. We also fixed the FWHM of the Lorentzian used to fit the telluric absorption to 3.5 pixels based upon our experience in analysing NIRSPEC data, while also to overcome the impacts of the poorly removed OH emission lines in the six frames at the beginning of WASP-52b's observations (section \ref{sec:dr}). We did this as the residual OH emission impacted \texttt{molecfit}'s ability to model the nearby H$_2$O absorption if the FWHM was allowed to vary (Figure \ref{fig:molecfit_model}). We note that \cite{Zhang2021} similarly fixed this parameter to 3.5 pixels in their \texttt{molecfit} modelling of NIRSPEC data. 

Given that order 71's wavelength coverage included fewer telluric absorption lines, we were not able to obtain a satisfactory fit to order 71 in isolation. Instead we found that using the best fitting parameters from order 70 (i.e., depth of the water column etc.) gave good fits when applied to order 71. 

Figure \ref{fig:molecfit_model} shows example WASP-52 and WASP-177 spectra before and after the telluric correction using \texttt{molecfit}. This also demonstrates the proximity of the \ion{He}{1} triplet to OH emission lines on both nights. However, aside from the first six frames for WAPS-52b, our A-B nod subtraction effectively removed these emission lines from our science spectra.

There appear to be some residual oscillations in the example spectrum of WASP-177 (Figure \ref{fig:molecfit_model}, bottom panel). One possibility behind these is residual fringing from our use of the Thin blocking filter (section \ref{sec:obs}), despite our fringing correction. However, the amplitudes of oscillations in the master spectra of WASP-177 (Figure \ref{fig:ss_w177}) are not significantly greater than those of WASP-52 (Figure \ref{fig:ss_w52}), for which we did not use the Thin blocking filter and thus avoided significant fringing. Therefore, we do not believe fringing is significantly affecting our results for WASP-177b.

Following the removal of telluric absorption from our spectra, we then shifted each of our spectra from the observer frame into the stellar rest frames and checked the accuracy of our wavelength solution. To shift our spectra into the stellar frame we corrected for the barycentric velocity, via \texttt{astropy}'s \citep{astropy:2013,astropy:2018} implementation of \cite{Wright2014}'s method, in addition to the systemic velocity and stellar reflex velocity caused by the close-in gas giant (using the parameters in Table \ref{tab:system_params}). To confirm our wavelength solutions, we cross-correlated our stellar spectra with Phoenix \citep{PHOENIX} model spectra for both stars. We found in both cases the wavelength solutions needed small corrections ($\sim 1$\,km\,s$^{-1}$). Despite these being small corrections ($\sim 1/3$ pixel) we applied them since we are sensitive to velocity shifts in the planets' \ion{He}{1} absorption of $\sim1$\,km\,s$^{-1}$ (section \ref{sec:trans_spec}).

At this stage, we had normalized, telluric-corrected stellar spectra in the stellar rest frame. However, we still needed to account for the poorly removed OH emission in WASP-52b's first six frames and cosmic rays in the spectra of both WASP-52 and WASP-177. To do this, we performed a sigma-replacement method \citep[e.g.,][]{Allart2017}. Specifically, we made a median-combined spectrum for both WASP-52 and WASP-177 and compared each spectral frame to the combined median. We then replaced any data points that deviated by $>4$ median absolute deviations from the combined median, with the corresponding median-combined data point. 

To avoid clipping out real planetary signal from our spectra, we masked the spectra within $\pm20$\,km\,s$^{-1}$ of the \ion{He}{1} triplet in the planets' rest frames. Instead, for outliers in the \ion{He}{1} triplet in the planets' rest frames, we removed whole frames from our analysis based on a fit to the planets' \ion{He}{1} light curves (see Appendix \ref{sec:frame_clipping}). This meant that we were neither clipping real signal nor being biased by outlying frames. By this method we removed frames 7 and 10 from order 70, and frames 8, 11, and 15 from order 71 for WASP-52 (10\% of our spectra). For WASP-177 we removed no frames from order 70 and frames 47 and 56 from order 71 (2\% of our spectra).

\section{Data analysis}
\label{sec:da}

\subsection{Creating the master spectra}

Our data analysis started with generating master in- and out-of-transit spectra so that we could obtain the in-transit excess absorption signal. The in- and out-of-transit spectra were constructed by taking the weighted mean of spectra that fell between the second and third contact points, and before and after the first and fourth contact points of the transit, respectively. These contact points were determined using the ephemerides of \cite{Mancini2017} for WASP-52b and \cite{Turner2019} for WASP-177b.

\begin{figure}
    \centering
    \includegraphics[scale=0.38]{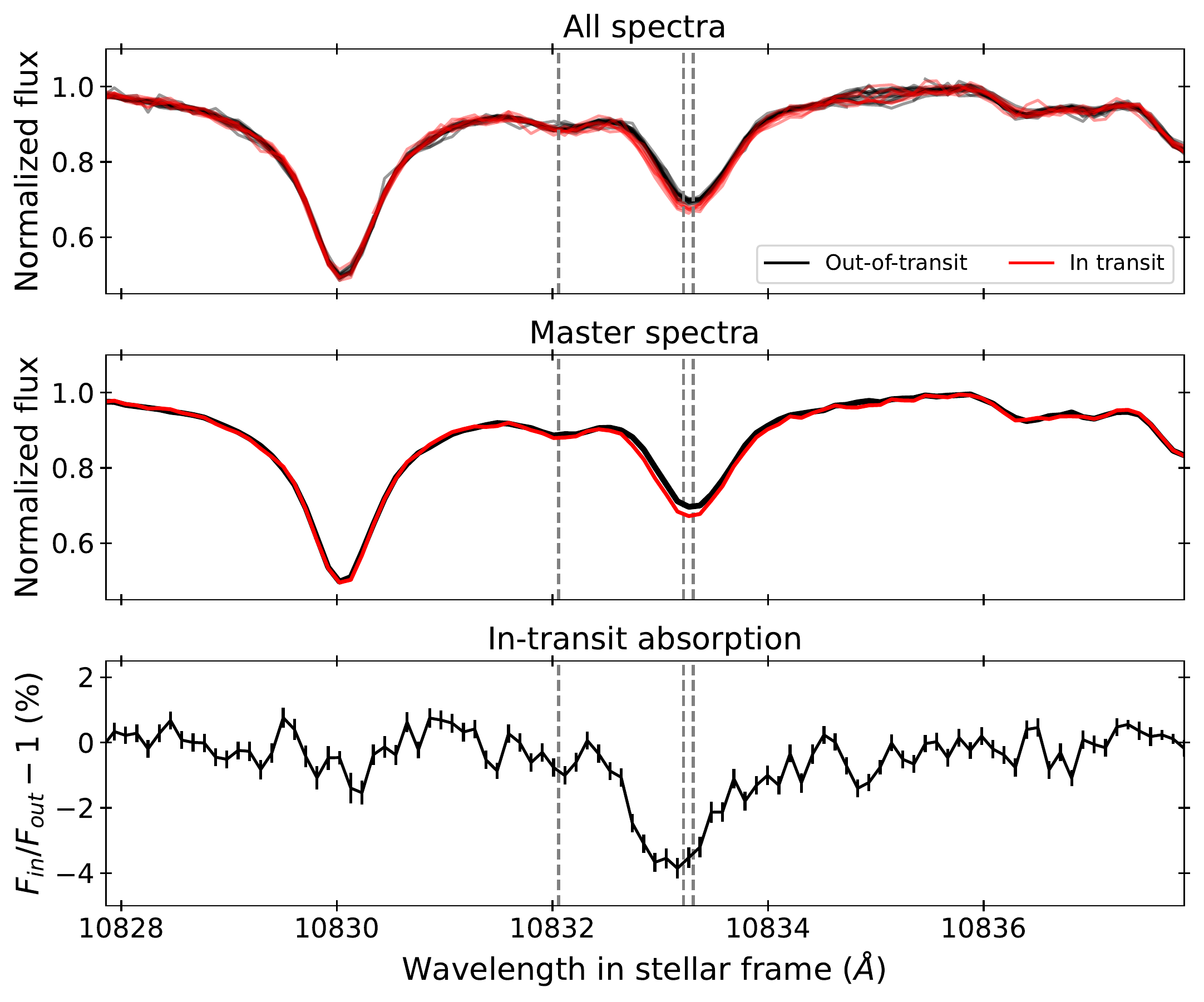}
   \caption{WASP-52's spectra centered on the helium triplet (shown by the vertical dashed lines). Top panel: the individual stellar spectra, showing the out-of-transit spectra (black) and the in-transit spectra (red). Middle panel: the combined, `master' out-of-transit and in-transit spectra. Bottom panel: the residual (excess) absorption centered on the helium triplet.}
    \label{fig:ss_w52}
\end{figure}

\begin{figure}
    \centering
    \includegraphics[scale=0.38]{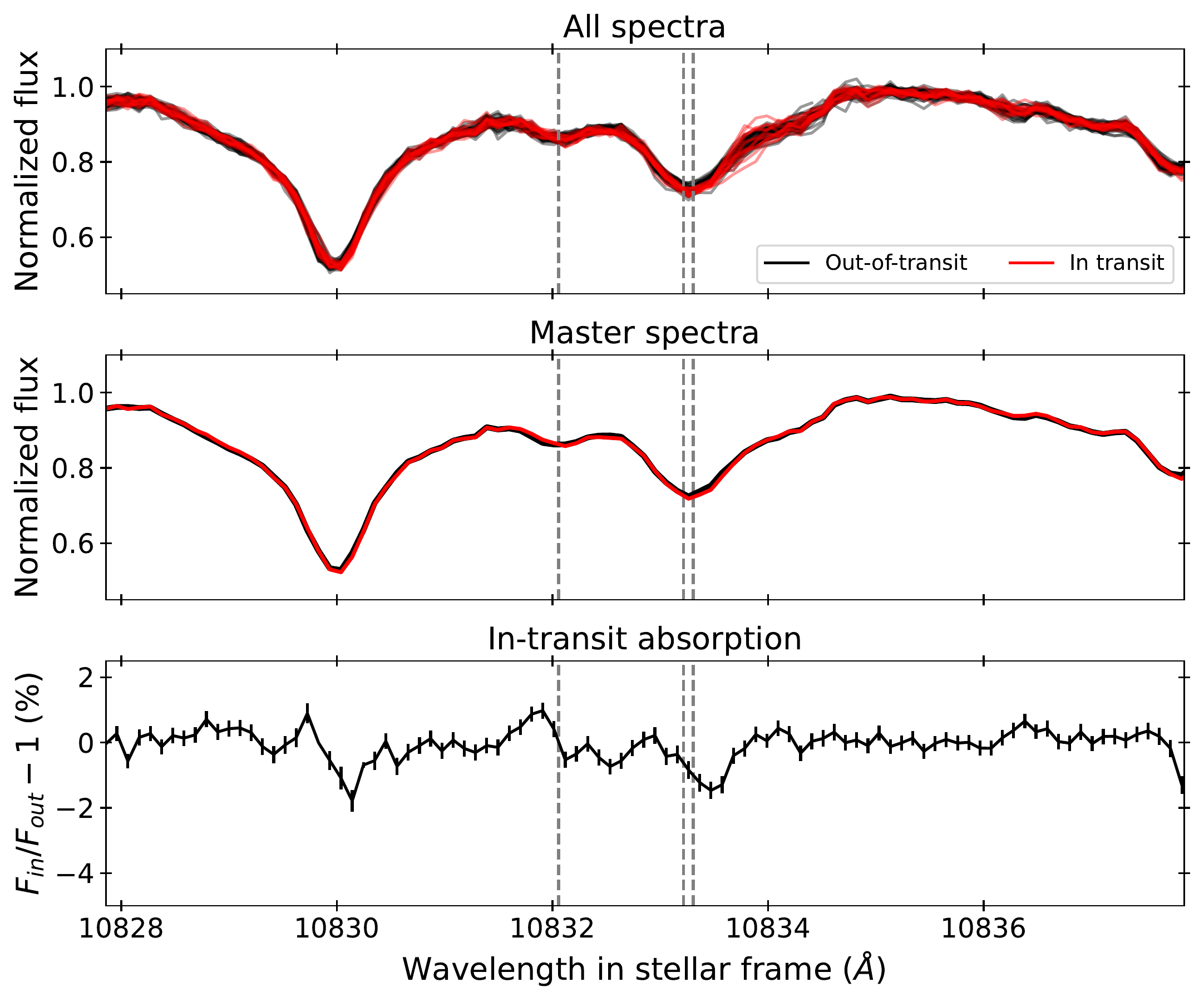}
    \caption{The same as Figure  \ref{fig:ss_w52} but for WASP-177.}
    \label{fig:ss_w177}
\end{figure}

 Figures \ref{fig:ss_w52} and \ref{fig:ss_w177} show the individual and master spectra for both WASP-52 and WASP-177. In these figures, we have combined orders 70 and 71 into a single spectrum. Figures \ref{fig:ss_w52} and \ref{fig:ss_w177}, demonstrate that there is excess absorption of $\sim4$\,\% centered on \ion{He}{1} for WASP-52b but no immediately apparent excess absorption for WASP-177b. We investigate this excess absorption in the following subsections.

\subsection{The phase-resolved absorption}

\begin{figure}
    \centering
    \includegraphics[scale=0.5]{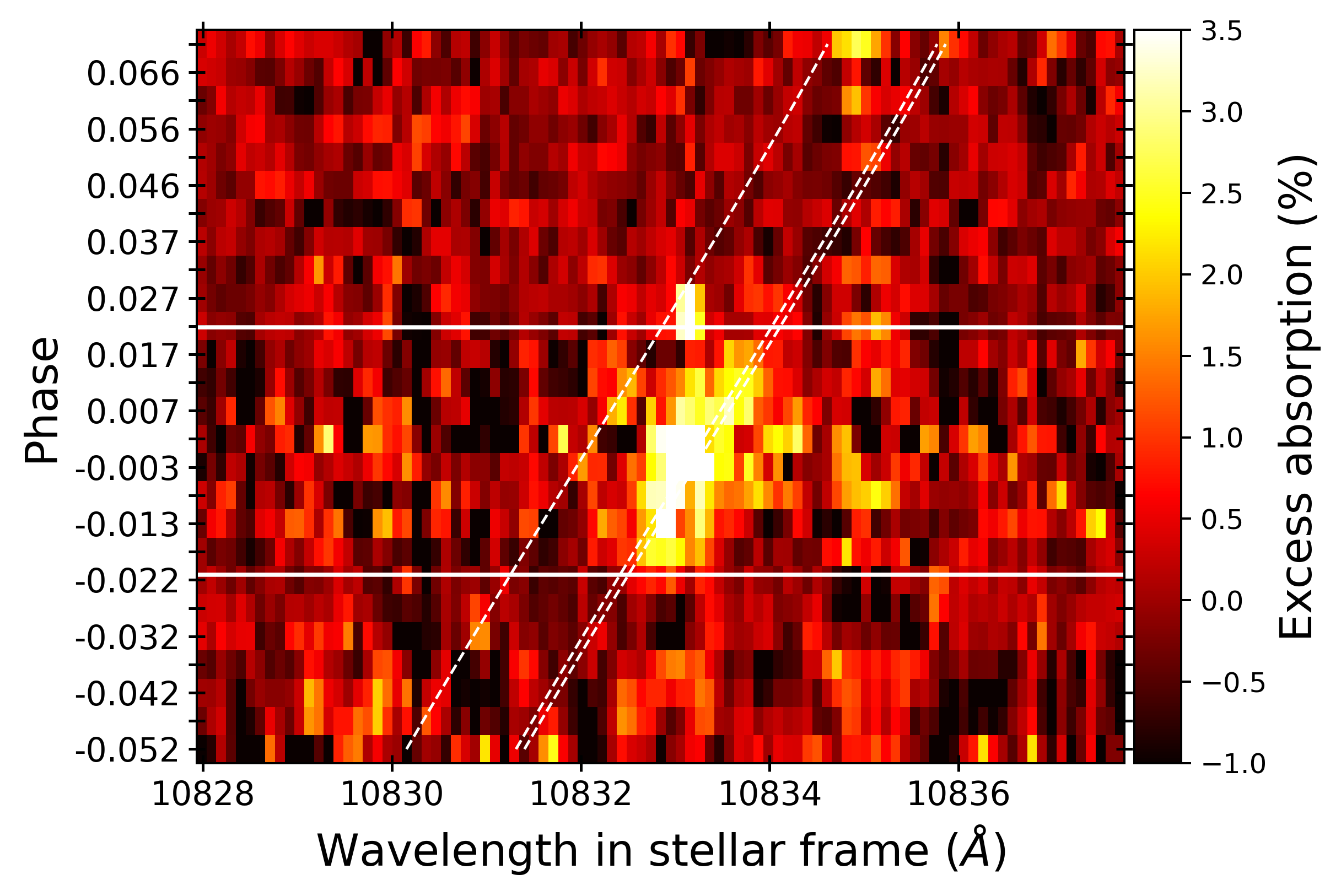}
    \caption{The phase-resolved excess absorption centered on the helium triplet during WASP-52b's transit. The wavelength is in the stellar rest frame. The horizontal white lines show the first and fourth contact points of WASP-52b's optical transit (using the ephemeris of \protect\citealp{Mancini2017}). The dashed white lines indicate the planet's orbital motion.}
    \label{fig:pra_w52}
\end{figure}

\begin{figure}
    \centering
    \includegraphics[scale=0.5]{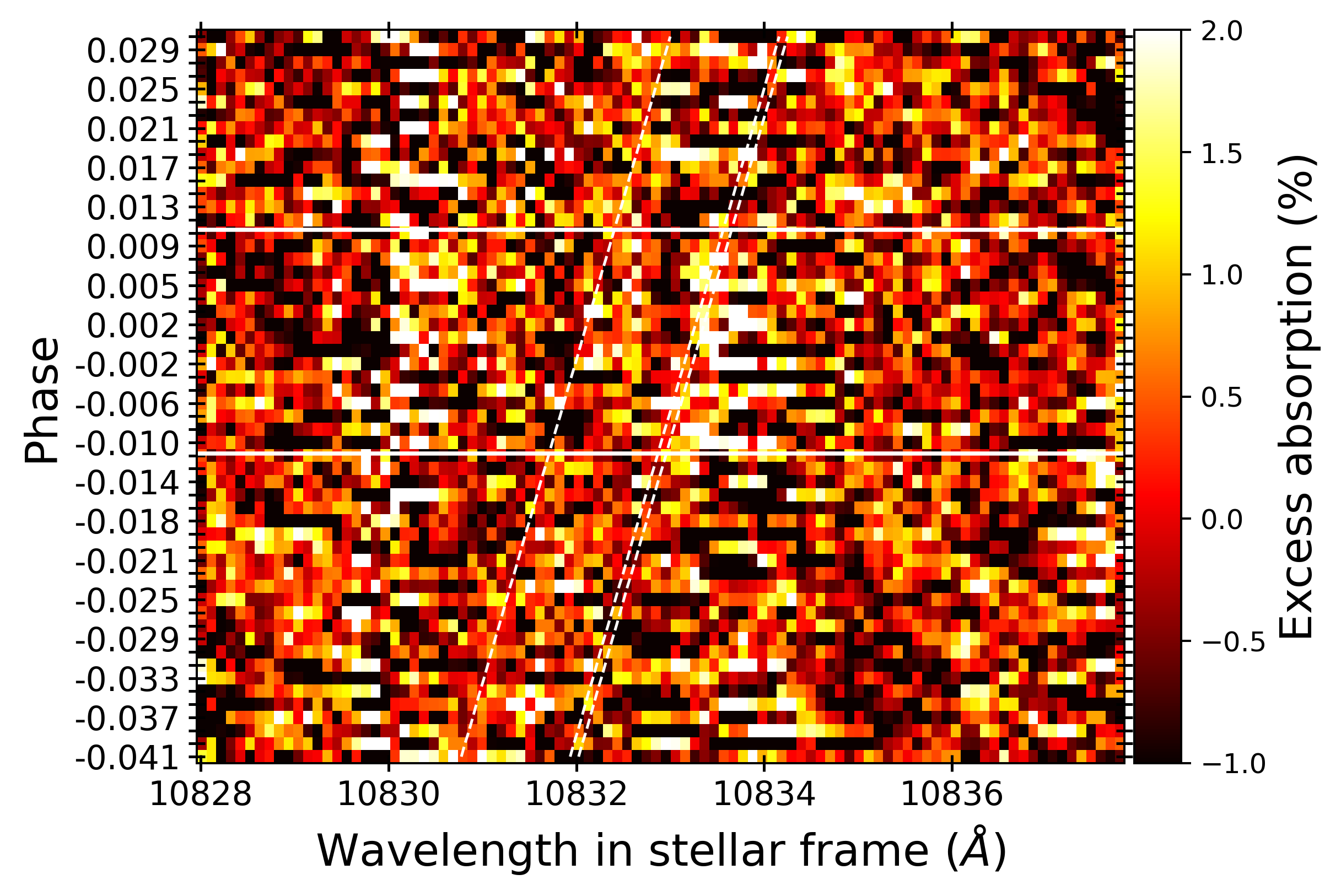}
    \caption{The phase-resolved excess absorption centered on the helium triplet during WASP-177b's transit, indicating no absorption visible by-eye. See Figure  \ref{fig:pra_w52} for details.}
    \label{fig:pra_w177}
\end{figure}

Figures \ref{fig:pra_w52} and \ref{fig:pra_w177} show the phase-resolved excess absorption for WASP-52b and WASP-177b. These figures show each spectral frame divided by the master out-of-transit spectrum. This was performed separately for order 70 and 71 for each planet. However, we then combined the residual spectra from both orders for our analysis. For WASP-52b (Figure  \ref{fig:pra_w52}), there is clear excess in-transit absorption while there is no significant phase-resolved absorption for WASP-177b (Figure  \ref{fig:pra_w177}).

\begin{figure}
    \centering
    \includegraphics[scale=0.5]{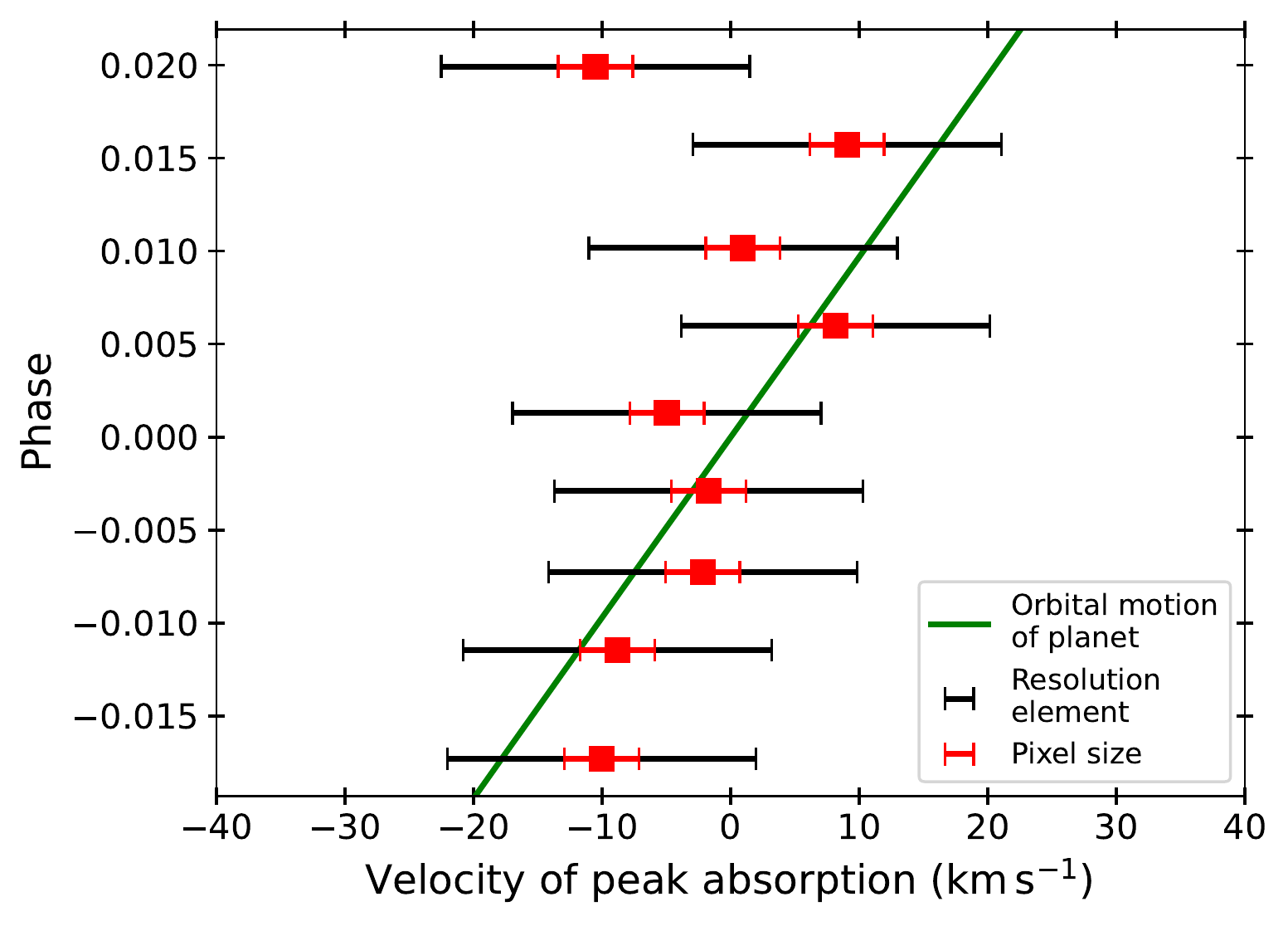}
    \caption{The velocity, in the stellar frame, of WASP-52b's peak \ion{He}{1} absorption during transit (red squares). The red error bars show the pixel size of NIRSPEC (0.11\,\AA, 3\,km\,s$^{-1}$) with the black error bars showing the resolution element (12\,km\,s$^{-1}$, $R = 25000$). The green line shows the planet's orbital motion given the parameters in Table \ref{tab:system_params}. We note that the final spectrum at phase 0.020 is at low signal to noise and so do not take this as evidence for blueshifted material.}
    \label{fig:aw_w52}
\end{figure}

Figure \ref{fig:aw_w52} shows the velocity, in the stellar frame, of WASP-52b's peak excess \ion{He}{1} absorption during its transit, calculated via fitting Gaussians to the \ion{He}{1} transmission spectrum in each frame. This demonstrates it is consistent with both the planet's orbital velocity and no velocity shift, when considering the resolution element of NIRSPEC (12\,km\,s$^{-1}$). We note that the final spectrum is at low signal to noise and so do not take this as evidence for blueshifted material.

\subsection{The transmission spectra}
\label{sec:trans_spec}

\begin{figure}
    \centering
    \includegraphics[scale=0.3]{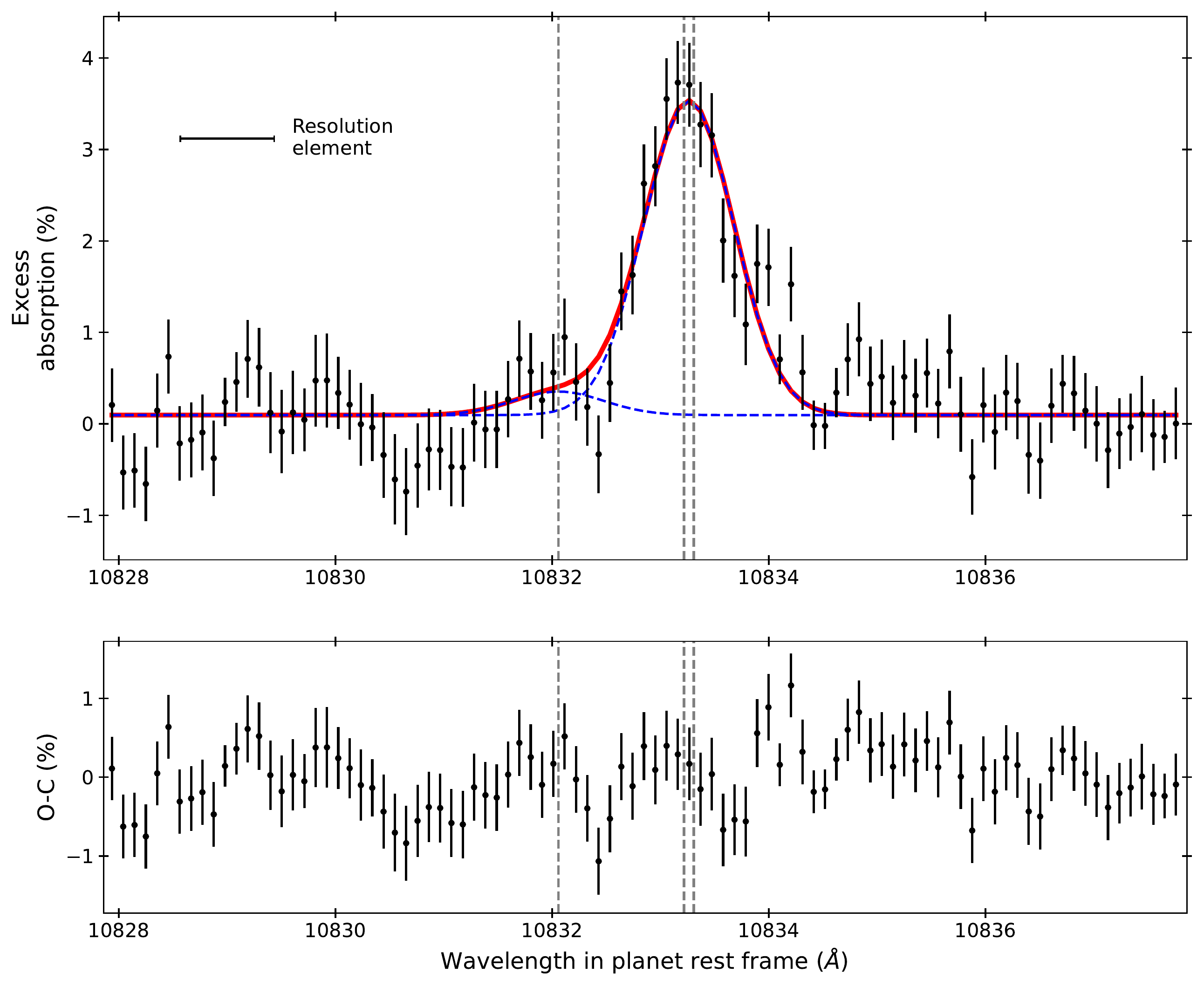}
    \caption{WASP-52b's transmission spectrum, centered on the helium triplet, whose location is shown by the vertical dashed gray lines. Top panel: the transmission spectrum (in units of \textit{excess} absorption) is shown by the black data points. The red line shows the fit of the summation of two Gaussians to the transmission spectrum. The blue dashed lines show the contribution of the two components of this double Gaussian. This reveals a peak amplitude of $3.44 \pm 0.31$\,\% ($11\sigma$) and no velocity shift ($0.00 \pm 1.19$\,km\,s$^{-1}$). Bottom panel: the residuals to the fit.}
    \label{fig:ts_w52}
\end{figure}

\begin{figure}
    \centering
    \includegraphics[scale=0.3]{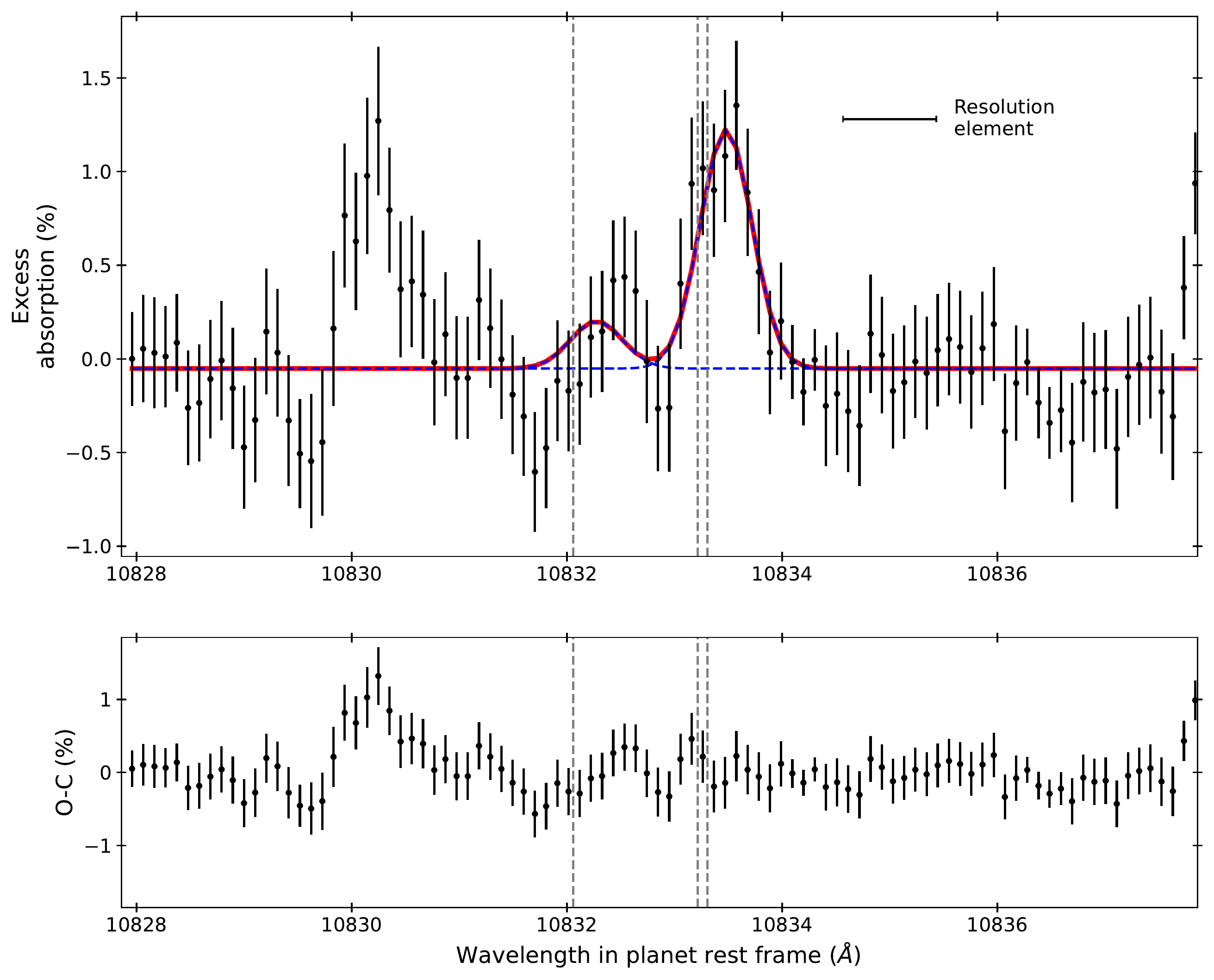}
    \caption{WASP-177b's transmission spectrum, centered on the helium triplet, which shows a peak at $1.28^{+0.30}_{-0.29}$\,\% that is redshifted by $+6.02 +/- 1.88$\,km\,s$^{-1}$. However, given the systematics in the spectrum we do not interpret this as strong evidence for helium.}
    \label{fig:ts_w177}
\end{figure}

By shifting the excess absorption to each planet's rest frame \citep[e.g.,][]{Wyttenbach2015,Wyttenbach2017,Allart2017}, we were able to construct the \ion{He}{1}  transmission spectra for both WASP-52b and WASP-177b. These are shown in Figures \ref{fig:ts_w52} and \ref{fig:ts_w177}. 

Similar to our treatment of WASP-107b's transmission spectrum in \cite{Kirk2020}, we initially analyzed the transmission spectra by fitting the summation of two Gaussians (which we refer to as a `double Gaussian') to quantify the excess absorption and wavelength shift. One Gaussian was centered on the weaker, bluer line at 10832.06\,\AA\ with the other centered on the two stronger, and blended, lines at 10833.22 and 10833.31\,\AA\ (vacuum wavelengths). We fitted for a  wavelength shift ($\Delta\lambda$) in the means of the two Gaussians to account for potential Doppler-shifted absorption. This wavelength shift was shared by both components of the Gaussian and was defined relative to the vacuum wavelengths of the helium triplet. The FWHM of the two Gaussians were set to be equal, given we expect the same instrumental and velocity broadening to apply to both components of the \ion{He}{1} absorption. The amplitudes of the two Gaussians ($A_1$ and $A_2$) were allowed to vary independently. We additionally fitted for a parameter ($C$) to account for imperfect normalization of the transmission spectrum, which effectively moved the double Gaussian up and down in $y$.

In total, the transmission spectrum was fitted with five parameters (FWHM, $\Delta\lambda$, $A_1$, $A_2$, and $C$). We used Markov chain Monte Carlo (MCMC) to explore the parameter space via the \texttt{emcee} Python package \citep{emcee}. We ran the MCMC with 42 walkers for 10000 steps each and discarded the first 5000 steps as burn-in. Following this initial run, we then rescaled the photometric uncertainties so that the best-fitting model gave a reduced $\chi^2=1$ to account for red noise not taken into account by the photometric uncertainties. We then ran a second MCMC with the same setup. 

For WASP-52b, we find the amplitude of the two Gaussians to be $0.26^{+0.24}_{-0.17}$ and $3.44 \pm 0.31$\,\% (11\,$\sigma$), respectively. We detect no velocity offset in WASP-52b's absorption ($\Delta v = 0.00 +/- 1.19$\,km\,s$^{-1}$).

For WASP-177b, the transmission spectrum shows evidence for excess absorption around the \ion{He}{1} triplet, however, at a lower amplitude than for WASP-52b, which is also consistent with the amplitude of systematic noise in the data (Figure \ref{fig:ts_w177}). Nevertheless, we also fitted WASP-177b's transmission spectrum with the same double Gaussian model, finding amplitudes of $0.25^{+0.23}_{-0.17}$\,\% and $1.28^{+0.30}_{-0.29}$\,\%. However, this absorption is redshifted by $+6.02 +/- 1.88$\,km\,s$^{-1}$. 

The systematic around the Si line in WASP-177b's transmission spectrum, and this apparent redshifted He absorption, may be due to imperfect wavelength calibration for this night. This is apparent when looking at the master-in and master-out spectra of WASP-177 (Figure \ref{fig:ss_w177}). Given the systematics in WASP-177b's transmission spectrum, we encourage additional observations to test the repeatability of this signal.

\subsection{The \ion{He}{1} light curves}
\label{sec:lightcurves}

Taking the residual spectra (stellar spectra divided by the master out-of-transit spectrum) in the planet rest frame, we generated light curves by integrating the residual flux in a bin centered on the \ion{He}{1} triplet. For the purposes of determining the transit depth as a function of bin width, we generated multiple light curves by varying the bin width from the resolution of NIRSPEC (0.43\,\AA\ $\approx 4$ pixels) to 15\,\AA\ in increments of 0.5\,\AA. For WASP-52b, we additionally created a light curve using a bin of width 6.35\,\AA\ to match the FWHM of the filter used by \cite{Vissapragada2020}. 

For both WASP-52b and WASP-177b, we fixed the planets' orbital periods, time of mid-transits, scaled semi-major axes, and inclinations to the values given in Table \ref{tab:system_params}. We fixed the quadratic limb darkening coefficients to values calculated by \texttt{LDTk} \citep{LDTK}, using the stellar parameters listed in Table \ref{tab:system_params}. We fitted only for $R_P/R_*$.

We used \texttt{batman} \citep{batman} to generate the light curves and fitted this using MCMC, again with \texttt{emcee} \citep{emcee}. In both cases we ran 20 walkers for 2000 steps each, discarding the first 1000 as burn in. We then again rescaled the photometric uncertainties to give a reduced $\chi^2 = 1$ and then ran the MCMC chains for a second time. 

\begin{figure*}
    \centering
    \includegraphics[scale=0.52]{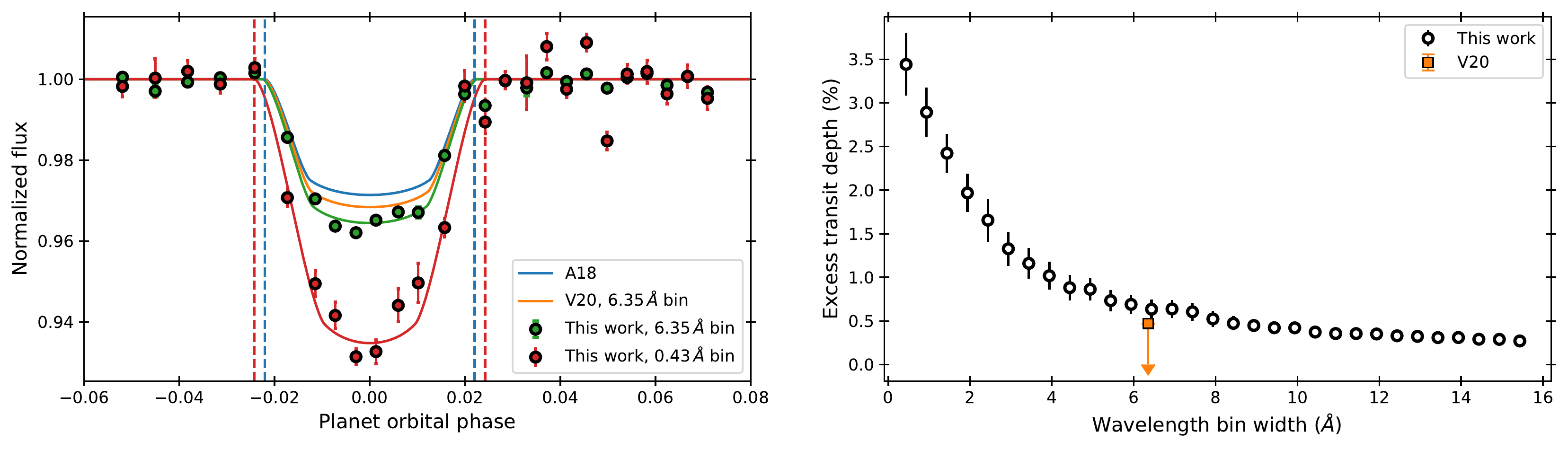}
    \caption{WASP-52b's transit light curve centered on the helium triplet. Left panel: the red points show the light curve found by integrating the residual absorption in a 0.43\,\AA-wide bin centered on the redder two lines of the \ion{He}{1} triplet. The red line shows a fit to these data that is then resampled to a finer time resolution. The green points show the light curve in a bin width matching the FWHM of the narrowband filter used by \protect\cite{Vissapragada2020}. The green line shows a fit to these data. The orange line shows the transit light curve observed by \protect\cite{Vissapragada2020} and the blue line shows the `continuum' optical light curve as measured by \protect\cite{Alam2018}. The red and blue vertical dashed lines indicate the first and fourth contact points of the transit models for the helium and continuum light curves, respectively. Right panel: the excess transit depth we observe with NIRSPEC as a function of the width of the bin that we integrate over. The orange square shows the 95th percentile upper limit on WASP-52b's excess \ion{He}{1} absorption measured by \protect\cite{Vissapragada2020} in a 6.35\,\AA-wide filter.}
    \label{fig:lc_w52}
\end{figure*}

\begin{figure*}
    \centering
    \includegraphics[scale=0.54]{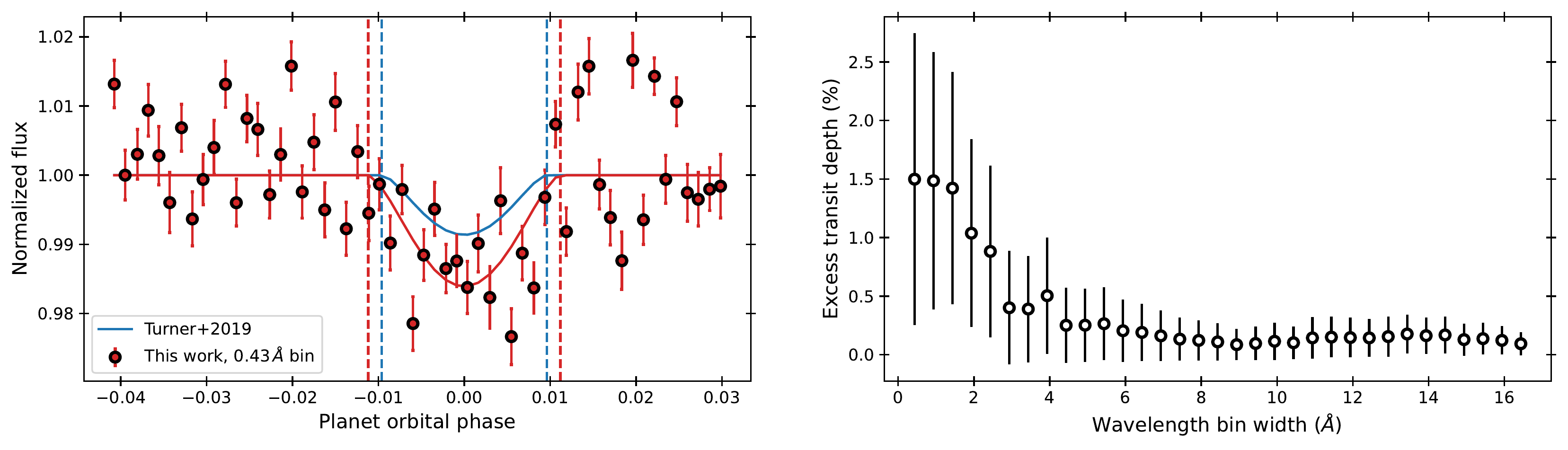}
    \caption{WASP-177b's transit light curve and the excess absorption we measure centered on the helium triplet, revealing no significant in-transit absorption. See Figure  \ref{fig:lc_w52} for details.}
    \label{fig:lc_w177}
\end{figure*}

Figures \ref{fig:lc_w52} and \ref{fig:lc_w177} show the light curves corresponding to the narrowest wavelength bins multiplied by the continuum light curves (from \citealp{Alam2018} for WASP-52b and \citealp{Turner2019} for WASP-177b). This multiplication is necessary to convert from \textit{excess} absorption to \textit{absolute} absorption. These figures also include the change in transit depth as a function of bin width. 

Fitting WASP-52b's helium light curve, we find that the excess transit depth is $3.44 \pm 0.36$\,\% in our narrowest bin, which is consistent with our transmission spectrum (Figure  \ref{fig:ts_w52}). The light curve is largely symmetric about the mid-point. To estimate the excess transit duration we observe, we resampled our fitted transit light curve (red line, Figure  \ref{fig:lc_w52}) to a time resolution of 30 seconds. Comparing this with the transit duration corresponding to \cite{Alam2018}'s optical light curve, we find that WASP-52b's transit duration is 11 minutes longer at the location of the \ion{He}{1} triplet in a 0.43\,\AA-wide bin. 

In our bin matching the filter used by \cite{Vissapragada2020} (green line, Figure  \ref{fig:lc_w52}), we measure excess absorption of $0.66 +/- 0.14$\,\% for WASP-52b. \cite{Vissapragada2020} place a 95th percentile upper limit on excess absorption in the helium bandpass of 0.47\,\%. Therefore in the same bandpass our result is 1.4\,$\sigma$ deeper than that of \cite{Vissapragada2020}.

JWST will be able to observe the \ion{He}{1} triplet with a maximum resolution of $R=2700$ with the G140H grism on the NIRSpec instrument, or equivalently $\Delta\lambda = 4$\,\AA. At this resolution, we predict excess \ion{He}{1} absorption of $\sim1$\,\% (Figure \ref{fig:lc_w52}) for WASP-52b. This should be readily detectable if NIRSpec can reach its predicted noise floor of $\sim20$\,ppm \cite[e.g.,][]{Greene2016,Batalha2017}.

For WASP-177b (Figure \ref{fig:lc_w177}), we detect no significant in-transit absorption from the \ion{He}{1} light curves.

\subsection{Bootstrap analysis}
\label{sec:bootstrap}

Following the approach of other ground-based high-resolution studies of narrow absorption lines \citep[e.g.,][]{Redfield2008,Salz2018,Alonso-Floriano2019}, we performed a bootstrap analysis as another check of the significance of our detection. For both orders 70 and 71, we randomly selected half of the in- and out-of-transit frames and calculated the median absorption in a 20\,km\,s$^{-1}$-wide bin centered on the two redder lines of the helium triplet. We repeated this process 5000 times. 

\begin{figure}
    \centering
    \includegraphics[scale=0.6]{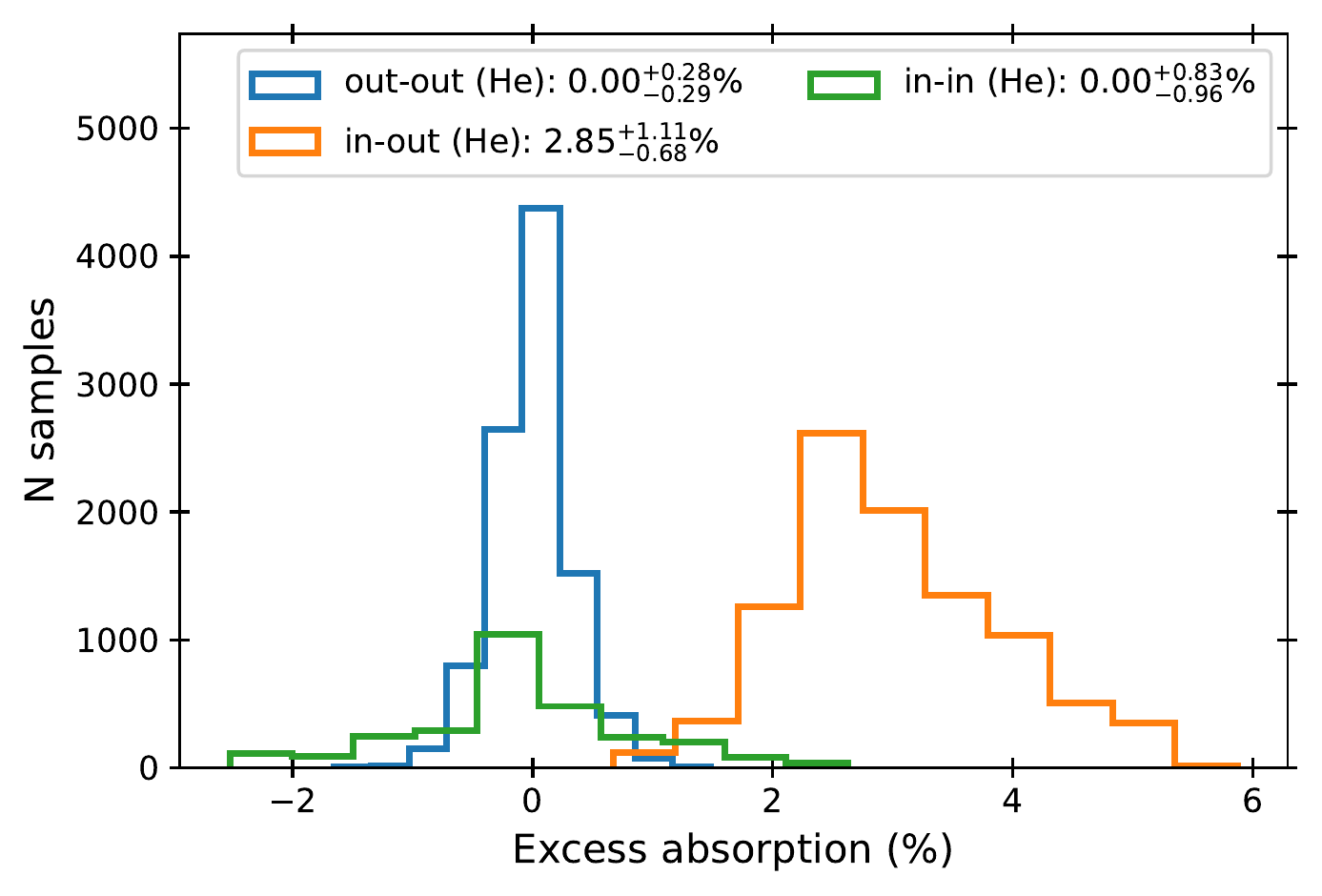}
    \caption{Our bootstrapping analysis of WASP-52b's \ion{He}{1} absorption. The out-out (blue), in-out (orange), and in-in (green) distributions for WASP-52b in a 20\,km\,s${^-1}$-wide bin centered on the helium triplet. For each distribution, the median, 16th and 84th percentiles are given in the legend.}
    \label{fig:bs_w52}
\end{figure}

Figure  \ref{fig:bs_w52} shows the results of this bootstrap analysis for WASP-52b, which reveals the in-minus-out distribution is $>0$ at $4.2\sigma$ confidence, while the out-minus-out distribution is centered on 0\,\%, as expected.

\begin{figure}
    \centering
    \includegraphics[scale=0.6]{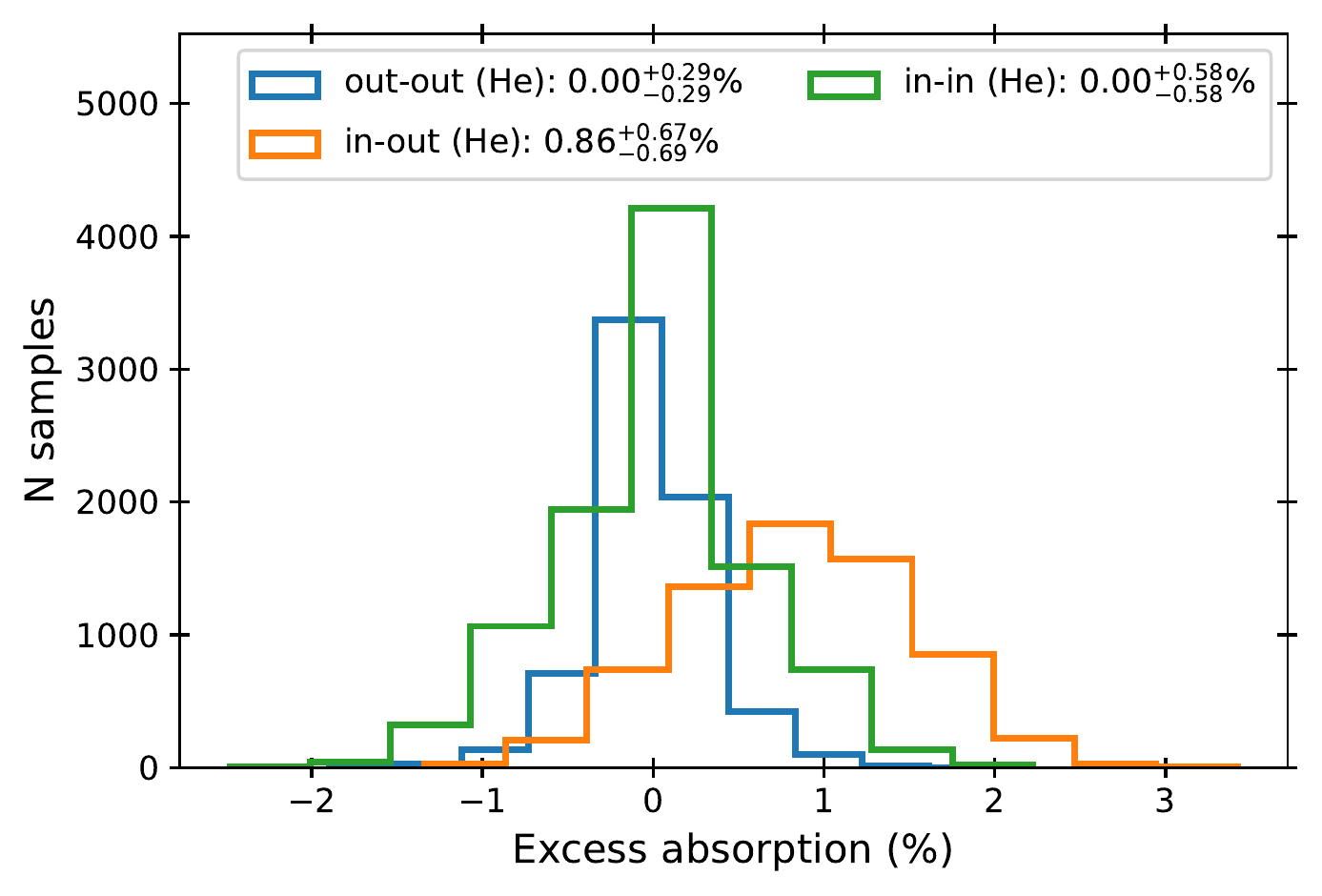}
    \caption{The distributions resulting from our bootstrap analysis for WASP-177b. See Figure  \ref{fig:bs_w52} for details.}
    \label{fig:bs_w177}
\end{figure}

Figure \ref{fig:bs_w177} shows the bootstrap analysis results for WASP-177b, revealing no significant excess in-transit absorption. We discuss this finding in the context of WASP-177b's transmission spectrum (Figure \ref{fig:ts_w177}) in section \ref{sec:discussion}.

\section{Atmospheric escape rate constraints and modelling}
\label{sec:1d_models}

We interpreted the metastable He transmission spectra of WASP-52b and WASP-177b using the one-dimensional atmospheric escape model {\tt p-winds\footnote{\url{https://github.com/ladsantos/p-winds}}} (version 1.2.3; \citealt{DosSantos2022}), which is based on the formulation presented in \cite{Oklopcic2018} and \cite{Lampon2020}, and has been benchmarked against the established EVE code \citep[e.g.,][]{Bourrier2013,Bourrier2015}. This model treats the escaping material as an isothermal Parker wind \citep{Parker1958} composed of only H+He, and finds the steady-state recombination/ionization solutions for the distribution of neutral H and He in the planetary upper atmosphere. The {\tt p-winds} code also solves the radiative transfer equation to determine the in-transit absorption caused by the planet and the escaping material. 

We fitted the co-added transmission spectra to an in-transit absorption model averaged in phase space. We used a nearby telluric absorption line to measure  the shape of the spectral point-spread function (PSF) of NIRSPEC near the helium triplet. Given our modeling of the telluric absorption in the spectra (section \ref{sec:pp}), we concluded the PSF is best represented by a Lorentzian with a full width at half maximum of $\sim 3.5$\,px. The implementation of {\tt p-winds}  takes into account both the temperature and the kinematic broadening caused by the planetary outflow. 

The two main free parameters of the model we fitted were the atmospheric escape rate $\dot{m}$ and the isothermal outflow temperature $T$. They were explored in log-space using {\tt emcee} with flat priors ($5 < \log{\dot{m}} < 15$\,g\,s$^{-1}$ and $3 < \log{T} < 5$\,K). The initial guess for the MCMC was obtained by performing a maximum likelihood estimation using the Nelder-Mead algorithm implemented in the {\tt optimize.minimize} function of SciPy \citep{scipy}. Another free parameter of the fit, for which we set no prior, was the bulk radial velocity shift $v_{\rm bulk}$ of the absorption in relation to the planetary rest frame.

We ran three different retrievals: Model 1) WASP-52b, with the H number fraction of the planetary outflow fixed to 0.90, and we explored the parameter space using 15 walkers and 7000 steps; Model 2) WASP-52b, with the H fraction set as a free parameter with a flat prior between 0.70 and 1.00, and we explored the parameter space using 20 walkers and 15000 steps; Model 3) WASP-177b, with a fixed H fraction of 0.90, 10 walkers and 7000 steps. The lower limit of the prior on H fraction was set semi-arbitrarily to avoid numerical errors that frequently occur at low H fractions. Recent studies that simultaneously fit both Lyman-$\alpha$ and metastable He absorption in HD~189733b and GJ~3470b show that their planetary outflows have H number fractions near 0.99 \citep{Lampon2021}. But whether this is a general trend among hot gas giants remains to be tested, and will likely require more observations. \citet{DosSantos2022} concluded that the retrieved atmospheric escape rate of HAT-P-11b \citep{Allart2018} is insensitive to the H fraction for values below $\sim$0.96 when using isothermal Parker wind models. For WASP-177b, we decided to not explore models where the H fraction is allowed to vary, since the detection is only tentative and we can only measure an upper limit for the atmospheric escape rate.

The relevant planetary parameters used in the fit are the same as shown in Table \ref{tab:system_params}. This modeling procedure requires knowledge about the EUV stellar spectrum \citep[e.g.,][]{Salz2018,Palle2020}. Since there is no such measurement for WASP-52 or WASP-177, we used the high-energy SED of eps~Eri for the first and HD\,40307 for the second, both measured by the MUSCLES survey \citep{France2016}. eps~Eri is the same spectral type as WASP-52 (K2V), with a similar age (eps~Eri: 0.4--1\,Gyr, \citealt{Mamajek2008,Baines2012}; WASP-52: $0.4^{+0.3}_{-0.2}$\,Gyr, \citealt{Hebrard2013}) and mass (eps~Eri: $0.82 \pm 0.06$\,M$_{\odot}$, \citealt{Baines2012}; WASP-52: $0.804 \pm 0.050$\,M$_{\odot}$, \citealt{Mancini2017}). For WASP-177, we chose to use the MUSCLES spectrum of HD\,40307, since it is an older ($\sim4.5$\,Gyr) K2.5 dwarf ($0.77 \pm 0.05$\,M$_{\odot}$, e.g., \citealt{Barnes2007,Sousa2008,Tuomi2013}) slightly closer to WASP-177 in age ($9.7 \pm 3.9$\,Gyr, $0.876 \pm 0.038$\,M$_{\odot}$, \citealt{Turner2019}). The stellar spectra were then scaled to the semi-major axis of WASP-52b and WASP-177b to reflect the amount of irradiation arriving at the top of the planetary atmosphere.

\begin{figure}
    \centering
    \includegraphics[scale=0.4]{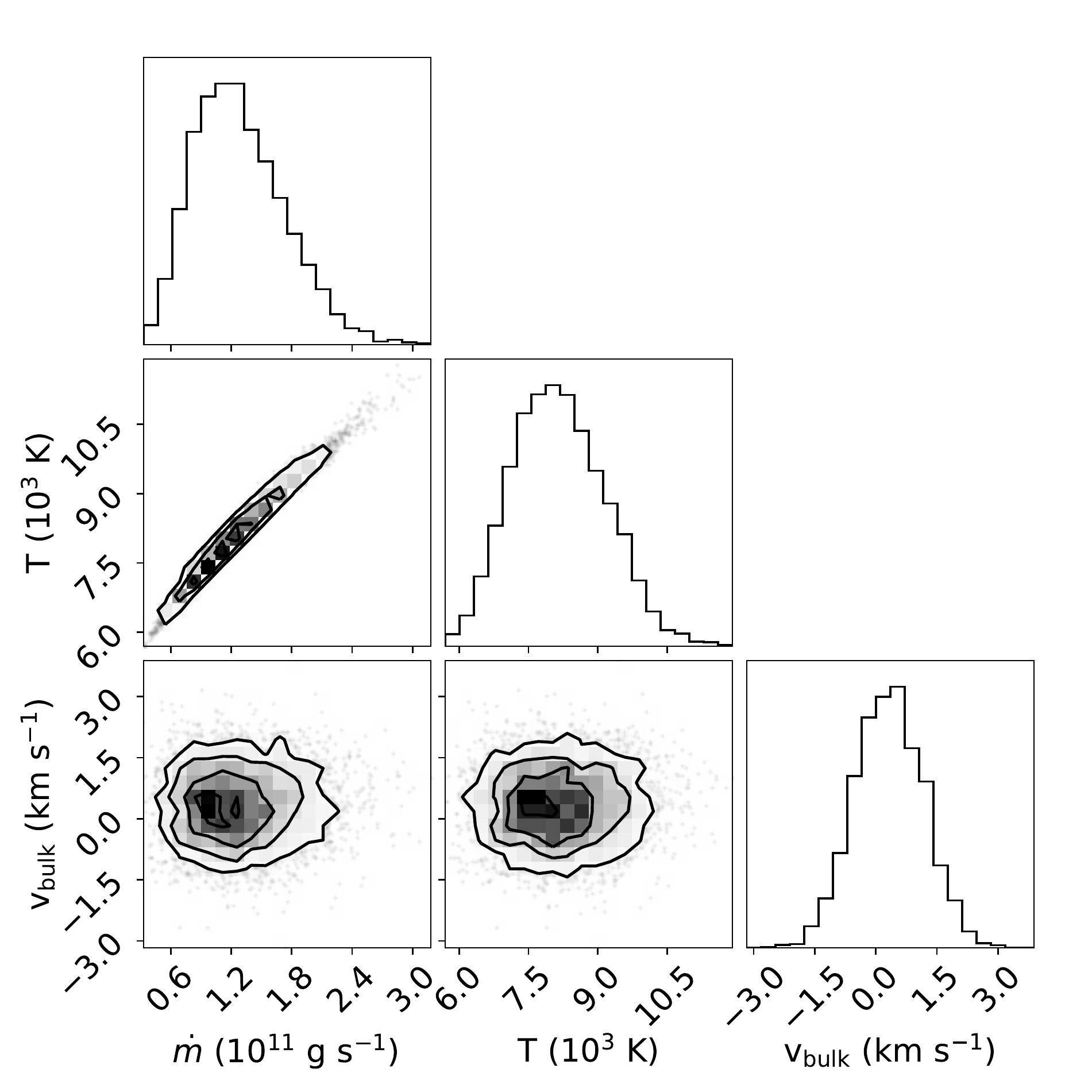}
    \caption{Posterior distributions of the atmospheric escape rate, outflow temperature, and bulk velocity of the outflow for WASP-52b. These results are based on one-dimensional, isothermal Parker wind models fitted to the observed transmission spectrum.}
    \label{fig:corner_3p_w52b}
\end{figure}

\begin{figure}
    \centering
    \includegraphics[scale=0.35]{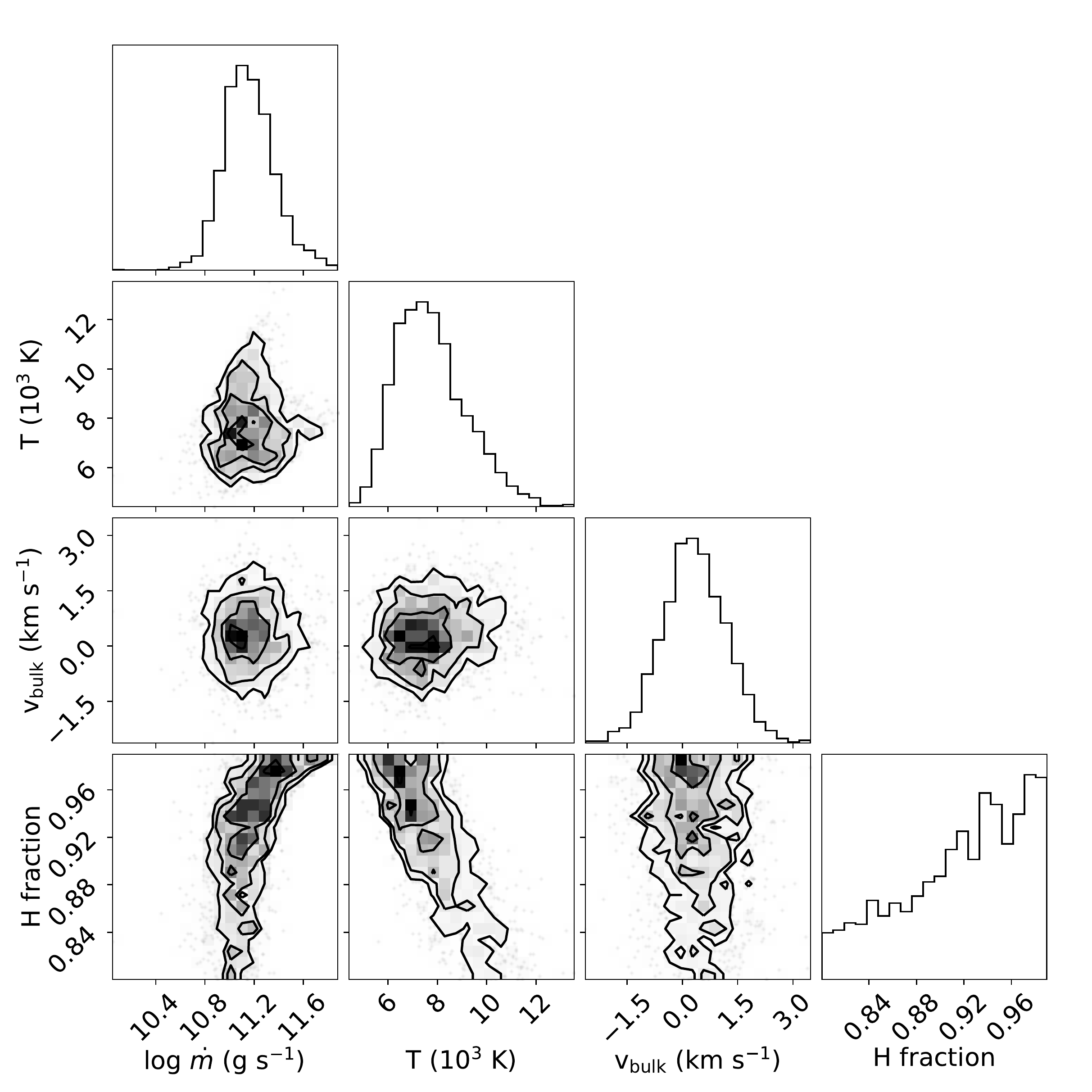}
    \caption{Same as Fig. \ref{fig:corner_3p_w52b}, but with H fraction as a free parameter. Note that the mass loss rate is represented in logarithmic scale.}
    \label{fig:corner_4p_w52b}
\end{figure}

\begin{figure}
    \centering
    \includegraphics[scale=0.35]{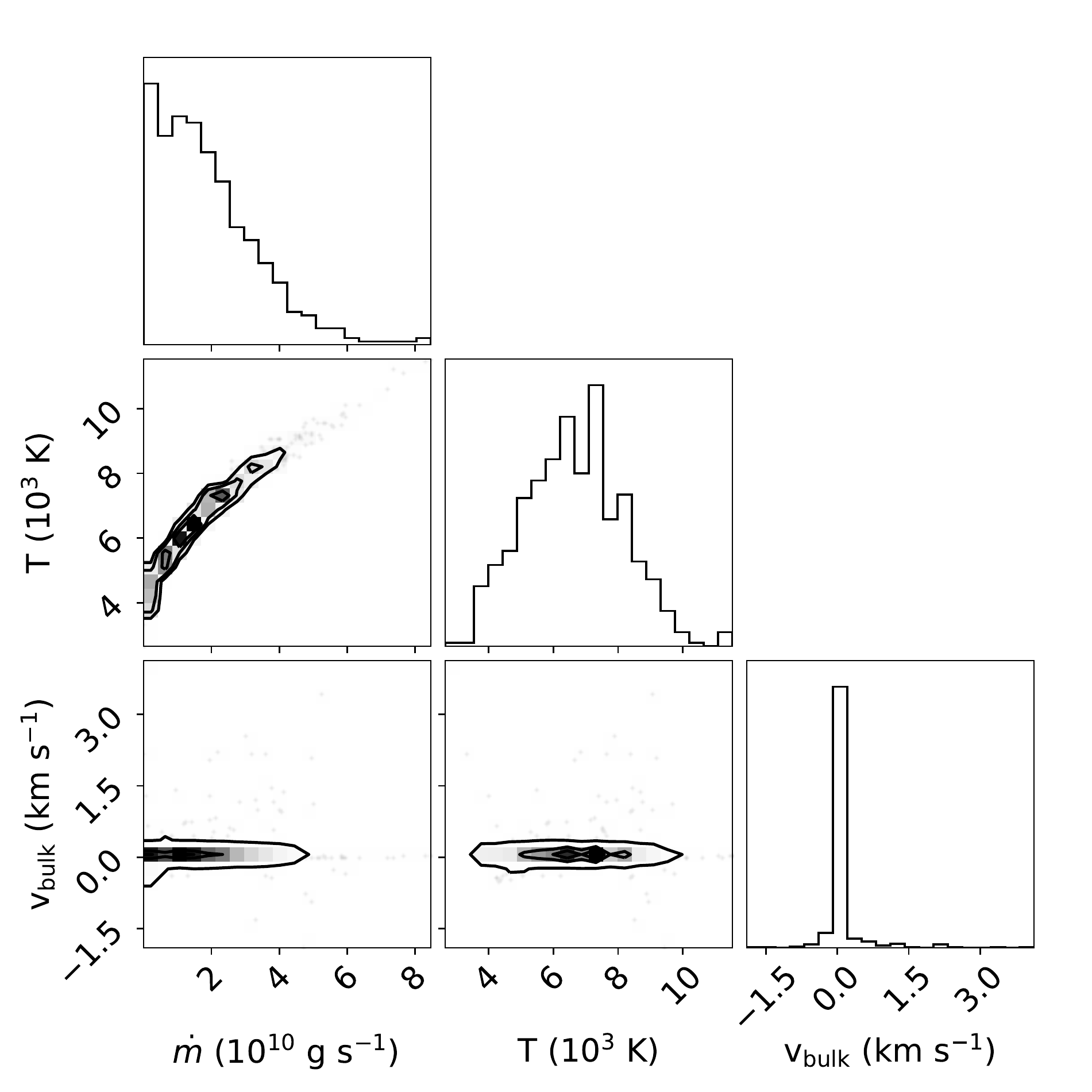}
    \caption{Same as Fig. \ref{fig:corner_3p_w52b}, but for WASP-177b}
    \label{fig:corner_3p_w177b}
\end{figure}

\begin{figure}
    \centering
    \includegraphics[scale=0.35]{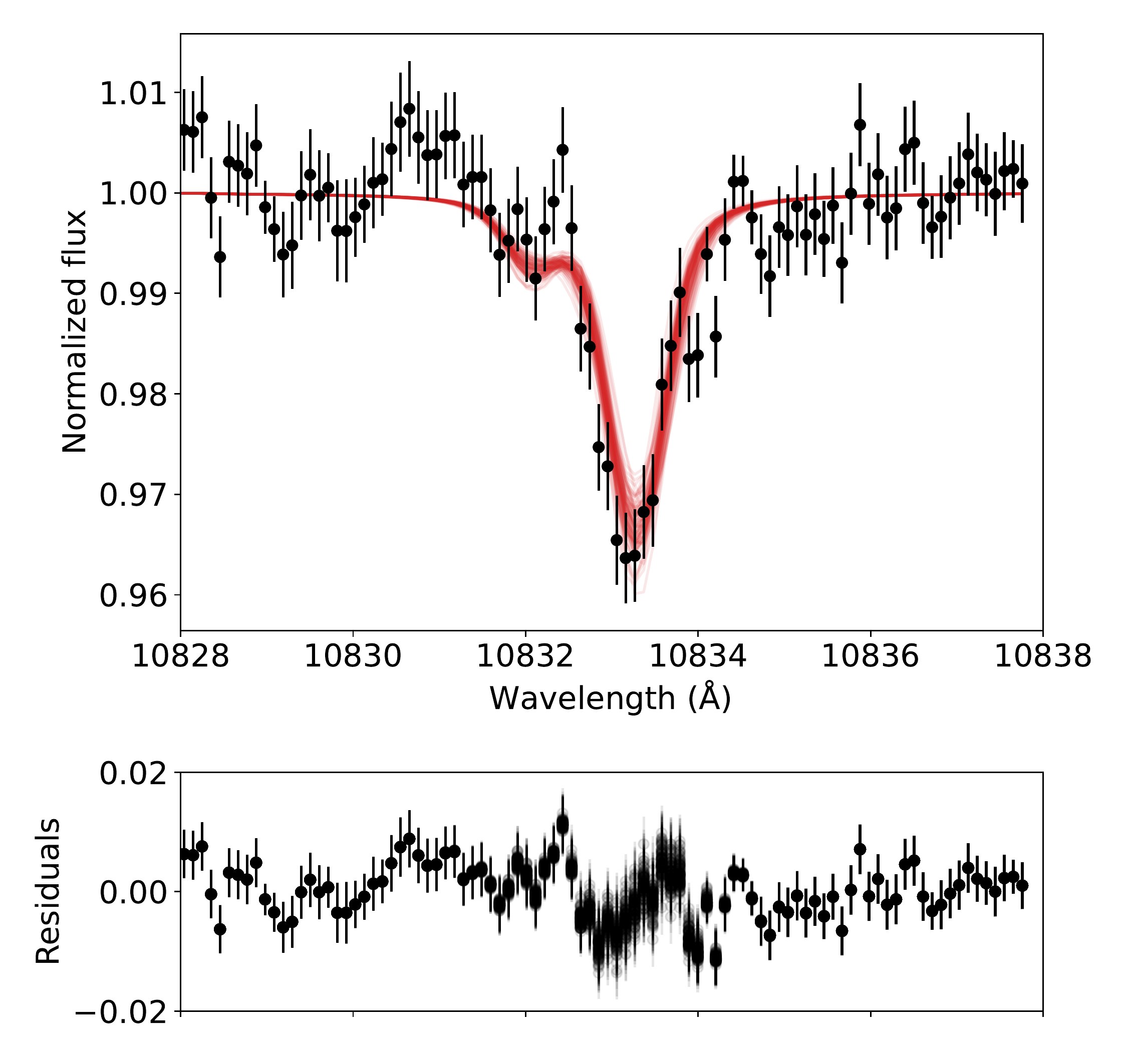}
    \caption{Sample of 100 one-dimensional, isothermal planetary wind models (red) with H fraction as a free parameter fitted to the observed transmission spectrum of WASP-52b (black symbols).}
    \label{fig:1dfit_4p_w52b}
\end{figure}

\begin{figure}
    \centering
    \includegraphics[scale=0.35]{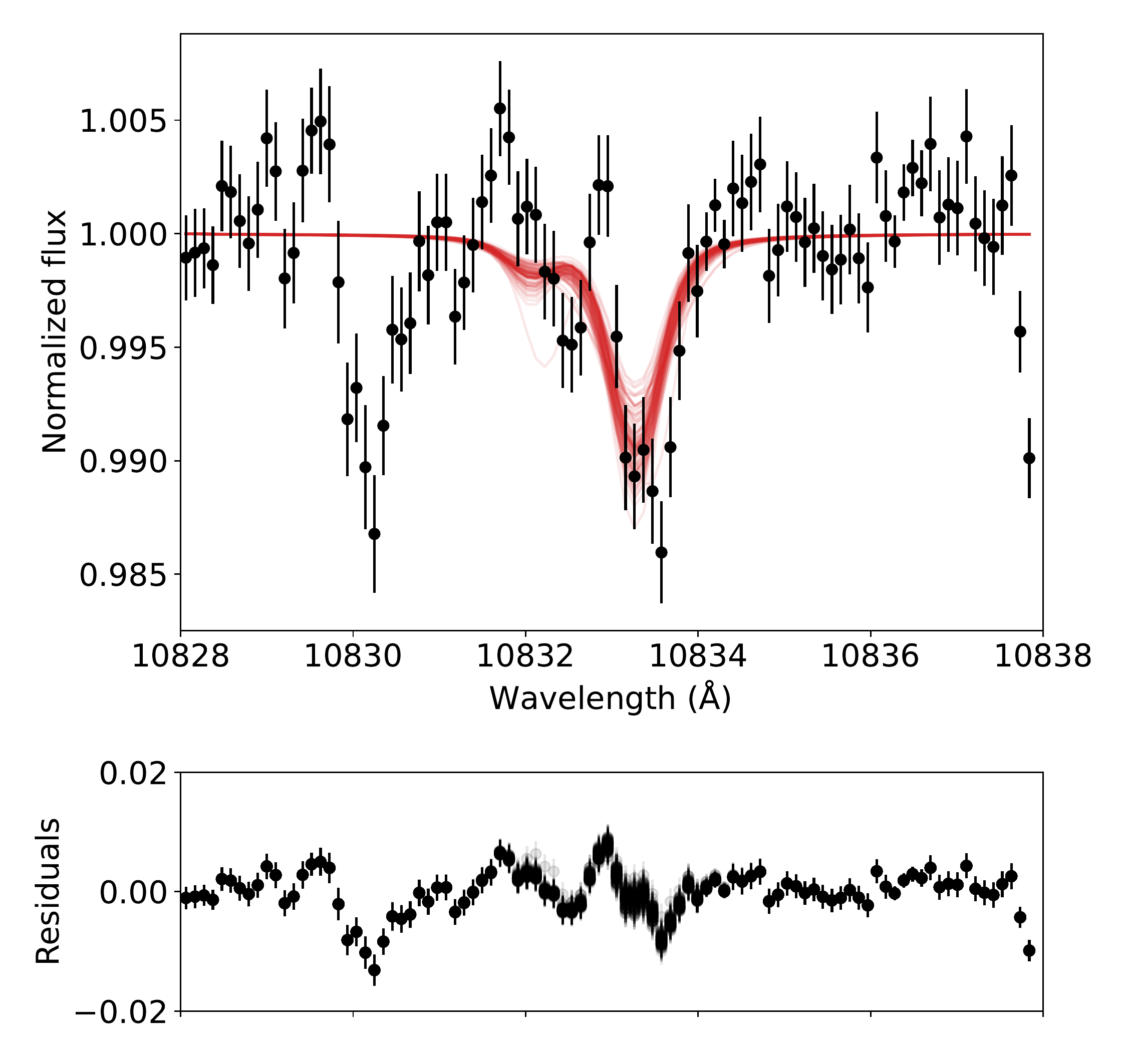}
    \caption{Same as Fig. \ref{fig:1dfit_4p_w52b}, but for WASP-177b and with the H fraction fixed to 0.90.}
    \label{fig:1dfit_3p_w177b}
\end{figure}

The results of the {\tt p-winds} fit to the observed transmission spectra of WASP-52b and WASP-177b are shown in Table \ref{tab:1d_results} and Figures \ref{fig:corner_3p_w52b}, \ref{fig:corner_4p_w52b}, and \ref{fig:corner_3p_w177b}. The resulting model transmission spectra in comparison with the observed data are shown in Figures \ref{fig:1dfit_4p_w52b} and \ref{fig:1dfit_3p_w177b}. We find that, independent of the H number fraction of the outflow, WASP-52b is most likely losing its atmosphere at a rate of $\sim1.4 \times 10^{11}$\,g\,s$^{-1}$, and that the temperature of the outflow is approximately $8000$\,K; the bulk radial velocity of the outflow in relation to the planetary rest frame is consistent with zero. Similar to the fit results for HAT-P-11b in \citet{DosSantos2022}, allowing the H fraction to vary as a free parameter did not significantly affect the retrieved $\dot{m}$ or $T$, except for increasing the uncertainties of the fit by a factor of $\sim$2. In the case of WASP-177b, we find a 3$\sigma$ upper limit of $7.9 \times 10^{10}$~g\,s$^{-1}$ for its mass loss rate, and an outflow temperature of approximately 6600\,K. 

For WASP-52b, the H number fraction is unconstrained at the 3$\sigma$ level. One important insight to be gained from the posteriors and correlation maps of Figure \ref{fig:corner_4p_w52b} is that the retrieved escape rate is mostly insensitive to H fractions below 0.90, above which value the retrieved $\dot{m}$ tends towards higher escape rates. There is an anti-correlation between the retrieved outflow temperature and the H fraction. These results mean that, if we are able to determine either the escape rate or outflow temperature independently of the metastable He transmission spectrum, the latter technique may be able to accurately determine the H fraction of the escaping atmosphere.

\begin{table*}[ht!]
\caption{1D modeling results for WASP-52b and WASP-177b.}
 \label{tab:1d_results}
    \centering
    \begin{tabular}{c|c|c|c|c} \hline
    & $\dot{m}$ & $T$ & $v_{\rm bulk}$ & H fraction \\
    & ($\times 10^{11}$\,g\,s$^{-1}$) & (K) & (km\,s$^{-1}$) & \\
    \hline     
    Model 1 (WASP-52b) & $1.2^{+0.5}_{-0.4}$ & $8100^{+1100}_{-900}$ & $+0.3 \pm 0.8$ & Fixed at $0.90$ \\
    Model 2 (WASP-52b) & $1.4^{+0.9}_{-0.5}$ & $7600^{+1600}_{-1200}$ & $+0.3 \pm 0.8$ & $>0.80$ (3-$\sigma$ confidence) \\
    Model 3 (WASP-177b) & $< 0.79$ (3-$\sigma$ confidence) & $6600 \pm 1500$ & $0.0 \pm 0.1$ & Fixed at $0.90$ \\
    \hline
    \end{tabular}
\end{table*}

\section{Discussion}
\label{sec:discussion}

\subsection{The \ion{He}{1} absorption of WASP-52b and WASP-177b}
\label{sec:discussion_he_abs}

As we showed in Section \ref{sec:trans_spec}, we measured significant ($11\sigma$) excess absorption by helium in WASP-52b's atmosphere and find tentative evidence for redshifted \ion{He}{1} absorption for WASP-177b, which is not confirmed by our light curve or bootstrap analysis.

For WASP-52b, we observe excess helium absorption of $3.44 \pm 0.31$\,\%. This excess absorption corresponds to $66 \pm 5$ atmospheric scale heights, where the scale height of the planet is 688\,km using the parameters in Table \ref{tab:system_params}. This in turn means that at the location of the helium triplet, and at the resolution of NIRSPEC, WASP-52b's excess helium absorption extends to $1.51 \pm 0.04$\,$\mathrm{R_P}$. Using the approximation of \cite{Eggleton1983} and the planet parameters given in Table \ref{tab:system_params}, we calculate WASP-52b's Roche radius to be 1.72\,$\mathrm{R_P}$. This means that the helium absorption we detect is close to filling the planet's Roche radius ($0.88+/-0.02 \times$ the Roche radius). Using 1D isothermal Parker wind models, we calculate WASP-52b's mass-loss rate to be $1.4 \times 10^{11}$\,g\,s$^{-1}$ or equivalently 0.5\,\% of its mass per Gyr. We discuss the possible consequences of 3D models in Section \ref{sec:3d_models}.

For WASP-177b, we see evidence for redshifted absorption ($\Delta v =  6.02 \pm 1.88$\,km\,s$^{-1}$) with an amplitude of $1.28^{+0.30}_{-0.29}$\,\% (equal to $23 \pm 5$\,$H$). However, the amplitude of this absorption is comparable to a systematic in the transmission spectrum associated with poor removal of the stellar Si line (Figure \ref{fig:ts_w177}), which we believe may be caused by imperfect wavelength calibration for WASP-177. This redshift amounts to approximately two pixels or half the resolution element.

Furthermore, our light curve (section \ref{sec:lightcurves}) and bootstrapping (section \ref{sec:bootstrap}) do not confirm any significant \ion{He}{1} absorption from WASP-177b. We therefore encourage additional observations of this planet to confirm or refute this possible hint of \ion{He}{1}.

If we instead place a $3\sigma$ upper limit on WASP-177b's \ion{He}{1} absorption based upon the standard deviation of its transmission spectrum (1.25\,\%), we find this is equal to an upper limit of 22\,$H$, where we calculate the scale height of the planet to be 872\,km using the parameters in Table \ref{tab:system_params}.

However, we also note that since the planet has a grazing transit (Figure  \ref{fig:lc_w177}), it is possible that these numbers are underestimated. Taking the $R_P/R_*$ ($0.1360^{+0.0129}_{-0.0052}$) and impact parameter ($b = 0.980^{+0.092}_{-0.060}$) of WASP-177b \citep{Turner2019}, we calculate that at mid-transit $55^{+21}_{-14}$\,\% of WASP-177b's atmosphere is being probed. Therefore, taking the $1\sigma$ lower bound on the amount of the planet's atmosphere that is being probed at mid-transit (41\,\%), and scaling our 22\,$H$ upper limit, the upper limit on WASP-177b's \ion{He}{1} absorption could be as high as 54\,$H$, assuming spherically symmetric \ion{He}{1} absorption.

\subsection{Stellar activity}

WASP-52 is an active star with numerous observations of magnetic activity regions occulted during transits of the planet \citep{Kirk2016,Louden2017,Mancini2017,Bruno2018,May2018}. Despite this activity, we interpret the helium absorption we detect as being planetary, not stellar, in nature. 

In a simulation study, \cite{Cauley2018} showed that the 10833\,\AA\ \ion{He}{1} triplet could be contaminated at the 0.1\,\% level in specific cases, but that these would likely lead to a \textit{dilution} of the signal, not an enhancement/spurious detection. This is significantly smaller than the $3.44 \pm 0.31$\,\% signal that we observe (Figure  \ref{fig:ts_w52}). Additionally, in \cite{Chen2020}'s study of WASP-52b's H-$\alpha$ absorption, the authors demonstrated that the $0.86 \pm 0.13$\,\% absorption they detected was not replicated in the activity indicator lines they used as a control sample. Finally, the good agreement between our study and that of \cite{Vissapragada2020} (Figure  \ref{fig:lc_w52}), for which the observation epochs were separated by a year, suggests non-variable planetary absorption. Taken together, we attribute the absorption we detect to WASP-52b's atmosphere not stellar activity. 

For WASP-177, \cite{Turner2019} attributed modulation in its photometry to active regions on the host star. However, given it is the same spectral type as WASP-52 but older ($9.7 \pm 3.9$\,Gyr as opposed to $0.4^{+0.3}_{-0.2}$\,Gyr, \citealt{Hebrard2013}), similar arguments apply and therefore we do not believe that our observations of WASP-177b are significantly impacted by activity.

\subsection{On the possible consequences of three-dimensional models of WASP-52b's atmospheric escape}
\label{sec:3d_models}

We have so far discussed inferences from spherical models of escaping planetary outflows. In reality, planetary winds escape in an orbiting frame and are shaped by the stellar wind environment of their host stars. 
Thus, the geometry of the escaped planetary material can be distorted by orbital effects and the interaction with the stellar wind \citep[e.g.][]{McCann2019, WangDai2021}. This can, in turn, affect the overall strength of the absorption signal and how it relates to the properties of the planetary outflow, such as the mass-loss rate.

These effects can only be fully studied in three dimensions with simulation models catered to a particular planet's parameters. Performing Bayesian inference with these sorts of models remains computationally intractable because of their expense. However, our use of 1D atmospheric profiles to infer the planetary mass-loss rate is justified by the fact that the both WASP-52b's helium absorption (Figure \ref{fig:ts_w52}) and light curve (Figure \ref{fig:lc_w52}) are symmetric, which also suggests we are probing the thermosphere and not the exosphere \citep[e.g., cf.\ Figure 4 of][]{Allart2019}.

Recent 3D simulations by \cite{MacLeodOklopcic2021} show that, in cases of relatively weak and moderate confinement of the planetary outflow by the stellar wind, the helium absorption originates from a region of unshocked planetary material which is not significantly affected by the interaction with the stellar wind  \citep[See Figure 2 of][]{MacLeodOklopcic2021}. As a result, the helium light curve has a high degree of symmetry around the transit midpoint, similar to what we see for WASP-52b (Figure \ref{fig:lc_w52}), and the absorption depth is consistent with the predictions of the spherically symmetric Parker wind models which do not include stellar winds at all. In the case of strong confinement by the stellar wind, the planetary outflow gets distorted, which results in a boosted absorption signal (compared to the 1D Parker wind model predictions) and an asymmetric light curve with a prolonged helium egress, i.e.\ a helium `tail'. 

Given the predicted $\sim1$\,\% amplitude of WASP-52b's \ion{He}{1} absorption at the resolution of JWST (section \ref{sec:lightcurves}), future observations with JWST could provide a more finely sampled light curve which is needed to fully assess the impact of stellar wind on the escaping material.

\subsection{On the potential correlation between \ion{He}{1} absorption and XUV irradiation}

Previous studies of exoplanetary helium absorption suggested evidence for a potential relation between XUV irradiation and the amplitude of \ion{He}{1} absorption observed for gas giant exoplanets \citep[e.g.,][]{Nortmann2018,Alonso-Floriano2019,dosSantos2020}. However, more recent results \citep{Casasayas-Barris2021,Fossati2021} are in disagreement with this tentative relation.

\begin{figure}
    \centering
    \includegraphics[scale=0.375]{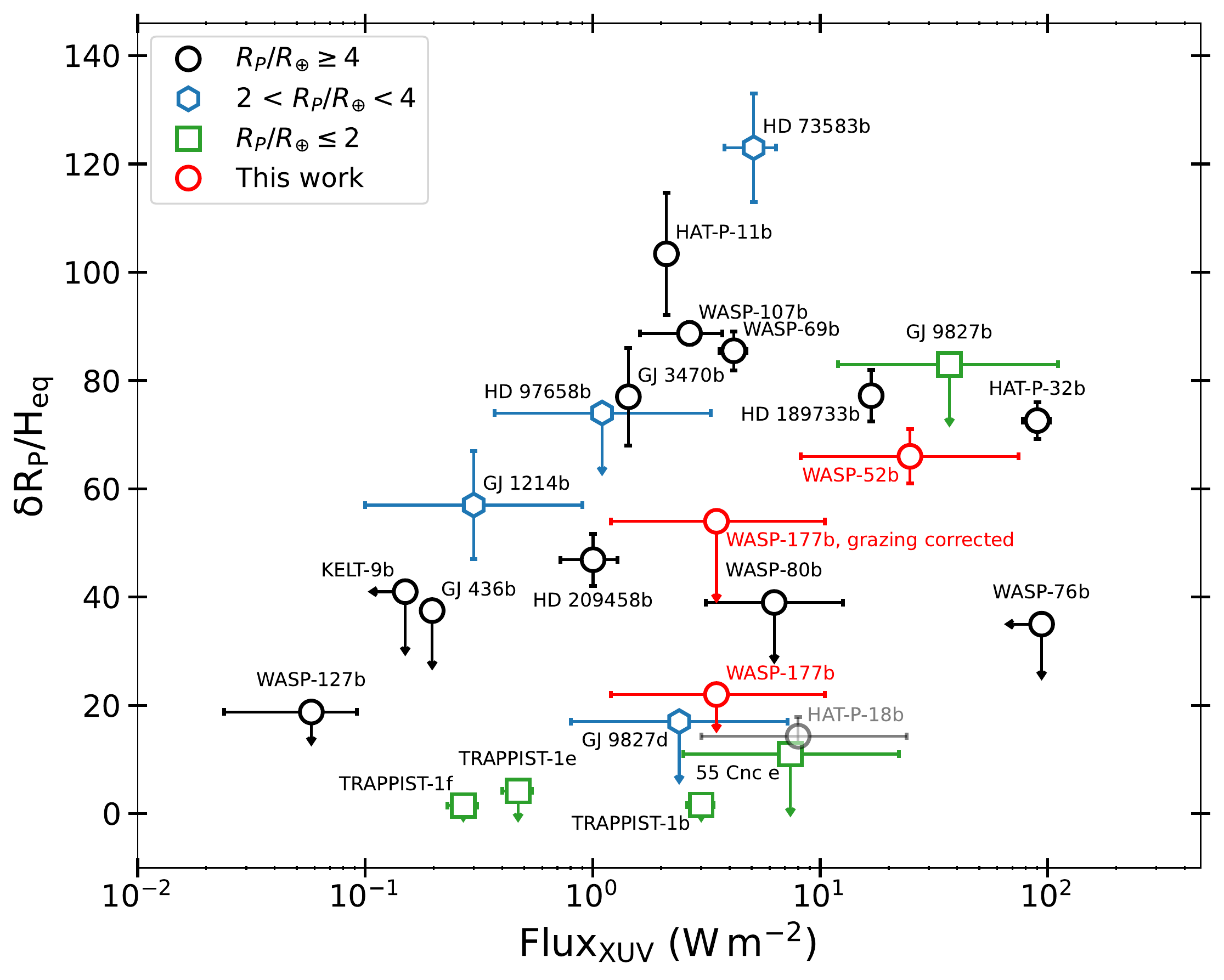}
    \caption{The published detections and non-detections/upper-limits of helium absorption in the literature, separated into gas giants (black circles), sub-Neptunes (blue hexagons) and super-Earths/Earths (green squares). These are plotted in terms of the XUV flux received against amplitude of excess helium absorption observed, in units of the planets' atmospheric scale heights. The red circles show our new results for WASP-52b and WASP-177b, including the upper limit after correcting for the planet's grazing transit configuration. We reiterate that WASP-52 and WASP-177 do not have measured XUV fluxes, but instead we use appropriate spectra from the MUSCLES survey \protect\citep{France2016} and assume these are accurate to within a factor of three (see text for details). We note that HAT-P-18b is shown with a lower opacity due to its detection with photometry \protect\citep{Paragas2021} which could be underestimating the full amplitude of this planet's absorption. The references for these planets are given in Table \ref{tab:lit_he}.}
    \label{fig:xuv_vs_delta_H}
\end{figure}

Figure \ref{fig:xuv_vs_delta_H} shows all exoplanets with well-constrained \ion{He}{1} absorption\footnote{All planets with detected \ion{He}{1} absorption or upper limits $< 100H$.}, along with our new findings for WASP-52b and WASP-177b. Following our 1D modeling (section \ref{sec:1d_models}), we adopted eps Eri for WASP-52 and HD\,40307 for WASP-177 to calculate the planets' XUV irradiation. Following \cite{Kasper2020} and \cite{Zhang2020}, we assume that our estimated XUV fluxes are accurate to within a factor of three, based on typical uncertainties in the reconstruction of stellar EUV fluxes \citep[e.g.,][]{Oklopcic2019}. However, this is likely an underestimation, since the gyrochronological and isochronal ages for WASP-52 and WASP-177 \citep{Hebrard2013,Mancini2017,Turner2019} lead to significantly different XUV fluxes when using empirical age--XUV luminosity relations \citep[e.g.,][]{Sanz-Forcada2011}. For the purposes of Figure \ref{fig:xuv_vs_delta_H}, we estimate $\mathrm{F_{XUV}} = 24.8^{+49.7}_{-16.6}$\,W\,m$^{-2}$ for WASP-52b and $\mathrm{F_{XUV}} = 3.5^{+7.0}_{-2.3}$\,W\,m$^{-2}$ for WASP-177b.

Our new results, taken together with recent results for WASP-76b \citep{Casasayas-Barris2021}, HAT-P-18b \citep{Paragas2021}, WASP-80b \citep{Fossati2021}, and HAT-P-32b \citep{Czesla2022}, suggest a shallower relation between XUV irradiation and \ion{He}{1}, if indeed such a relation exists. However, it is important to consider that WASP-177b's transit is grazing, and so its amplitude may be as large as 54\,$H$ (section \ref{sec:discussion_he_abs}), while HAT-P-18b's detection resulted from a narrowband filter which may also be underestimating the full amplitude of its \ion{He}{1} absorption. Futhermore, the observations of WASP-76b were hampered by telluric absorption \citep{Casasayas-Barris2021}. Therefore additional observations are needed to test the existence of such a relation.

While this manuscript was under review, \cite{Poppenhaeger2022} published a subset of literature \ion{He}{1} results, finding that the amplitude of exoplanetary \ion{He}{1} absorption is more strongly correlated to narrow-band EUV fluxes that take into account the stellar coronal iron abundances. We will look for a similar correlation in the updated sample of \ion{He}{1}-targeted exoplanets in a future work.

\subsection{Helium studies and the Neptune desert}

The Neptune desert is the name given to the observed dearth of short-period Neptunes in the exoplanet population \citep[e.g.,][]{Mazeh2016}. It has been suggested that this is the result of atmospheric loss; planets that initially fell within this desert were quickly stripped of their atmospheres and subsequently migrated out of the desert toward smaller masses and radii \citep[e.g.,][]{Kurokawa2014,Matsakos2016,Owen2018,Allan2019,Hallatt2021}. 

\begin{figure}
    \centering
    \includegraphics[scale=0.325]{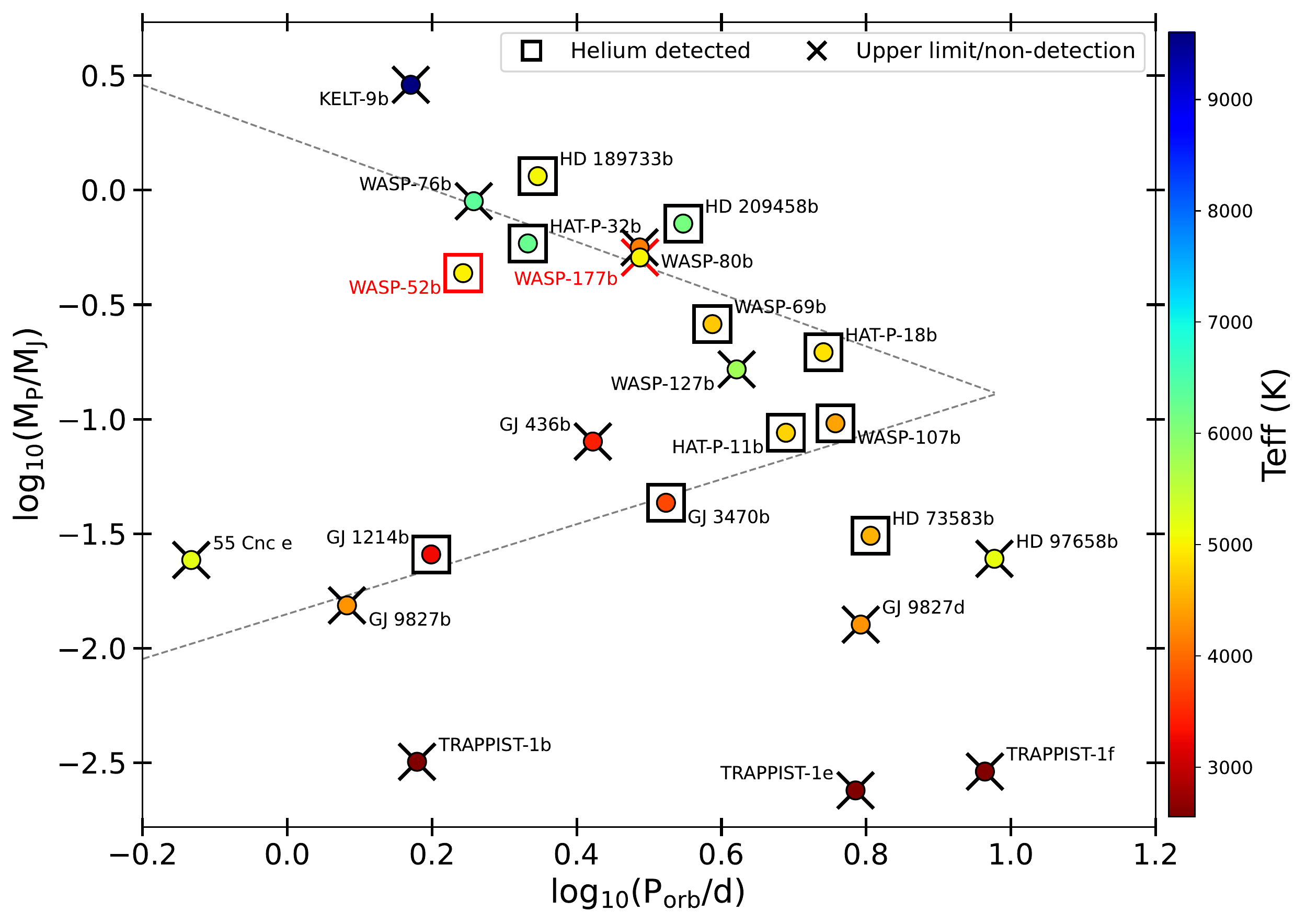}
    \caption{The sample of exoplanets that have been the focus of helium studies, along with the boundaries of the Neptune desert as defined by \protect\cite{Mazeh2016} (dashed gray lines). Each data point's color is related to the effective temperature of its host star. Those exoplanets with detections of helium are shown within squares and those with upper limits/non-detections are shown with crosses. The sizes of the uncertainties are smaller than the plot symbols. WASP-52b and WASP-177b (this work) are shown in red. See Table \protect\ref{tab:lit_he} for references.}
    \label{fig:ND}
\end{figure}

Given the rapid increase in the number of exoplanets that have been the focus of published helium observations, we can start to interpret these in the context of the Neptune desert. Figure \ref{fig:ND} shows the sample of published exoplanetary helium observations (Table \ref{tab:lit_he}) along with the boundaries of the Neptune desert as defined by \cite{Mazeh2016}. This figure reiterates the finding of \cite{Oklopcic2019} that K stars are the most favorable for studies of helium as most exoplanets with helium detections orbit stars with $T_{\mathrm{eff}} \approx 5000$\,K. 

If atmospheric loss is responsible for the Neptune desert, we might expect planets falling within the boundaries of the desert to be losing their atmospheres. Figure \ref{fig:ND} shows that several exoplanets with non-detections of helium absorption reside within the boundaries of the desert. However, since these planets do not orbit K stars, it is possible that they are losing their atmospheres but helium is not a sensitive probe.

Considering only WASP-52b and WASP-177b on Figure \ref{fig:ND}, we see that WASP-52b sits inside the Neptune desert and is losing its atmosphere at a significant rate. WASP-177b sits at the edge of the desert and due to systematics in our data and its grazing transit, we cannot say with confidence whether the planet is or is not losing its atmosphere. Nevertheless, Figure \ref{fig:ND} demonstrates the potential of the 10,830\,\AA\ \ion{He}{1} triplet to probe the origins of the Neptune desert, which motivates further observations of exoplanets in this parameter space.

\section{Conclusions}
\label{sec:conclusions}

In this study, we used the NIRSPEC instrument on the Keck II telescope to search for helium at 10833\,\AA\ in the atmospheres of the inflated hot gas giants WASP-52b and WASP-177b, both of which orbit K-type stars.

We detect significant excess absorption by helium in the atmosphere of WASP-52b, with an amplitude of $3.44 \pm 0.31$\,\% (11\,$\sigma$), or equivalently, $66 \pm 5$ atmospheric scale heights that is centered in the planet's rest frame ($\Delta v = 0.00 \pm 1.19$\,km\,s$^{-1}$). This absorption amplitude means that the planet is close to filling its Roche lobe. Using 1D isothermal Parker wind models, we find that WASP-52b is losing its mass at a rate of $\sim 1.4 \times 10^{11}$\,g\,s$^{-1}$, or equivalently, 0.5\,\% of its mass per Gyr. This is the first high-resolution detection of WASP-52b's escaping atmosphere. 

For WASP-177b, we find evidence for helium-like absorption of $1.28^{+0.30}_{-0.29}$\,\% in the planet's transmission spectrum. However, its anomalous redshift ($\Delta v = +6.02 \pm 1.88$\,km\,s$^{-1}$) combined with a lack of confirmation from light curve and bootstrap analyses means we do not interpret this as significant evidence for a detection of \ion{He}{1} in the planet's atmosphere. We therefore place a  $3\sigma$ upper limit on the planet's absorption of 1.25\,\%, or equivalently 22 atmospheric scale heights. However, because of the planet's grazing transit we may be underestimating the true extent of its helium absorption, which could be as much as 54 scale heights. Our 1D modelling of WASP-177b's helium transmission spectrum places a $3\sigma$ upper limit on the planet's escape rate of $7.9 \times 10^{10}$\,g\,s$^{-1}$.

Our results, taken together with recent results in the literature, raise doubts about the existence of a relation between XUV irradiation and \ion{He}{1} amplitude. Nevertheless, our results highlight the important role that \ion{He}{1} can play in understanding exoplanet atmosphere escape and how it impacts the exoplanet population through features like the Neptune desert.

\section*{Acknowledgements}
The data presented herein were obtained at the W.\ M.\ Keck Observatory, which is operated as a scientific partnership among the California Institute of Technology, the University of California and the National Aeronautics and Space Administration. The Observatory was made possible by the generous financial support of the W. M. Keck Foundation. The authors wish to recognize and acknowledge the very significant cultural role and reverence that the summit of Maunakea has always had within the indigenous Hawaiian community.  We are most fortunate to have the opportunity to conduct observations from this mountain.

We are grateful to our support astronomer Greg Doppmann for his guidance during the observations and data reduction. LADS acknowledges financial support from the European Research Council (ERC) under the European Union's Horizon 2020 research and innovation program, project {\sc Four Aces} (grant agreement No 724427); this work has partially been carried out in the frame of the National Centre for Competence in Research PlanetS supported by the Swiss National Science Foundation (SNSF). LZ would like to thank the support by the U.S. Department of Energy under awards DE-NA0003904 and DE-FOA0002633 (to S.B.J., principal investigator, and collaborator Li Zeng) with Harvard University and by the Sandia Z Fundamental Science Program. This research represents the authors’ views and not those of the Department of Energy.  Finally, we also thank the anonymous referee for their comments which led to an improvement in the quality of the manuscript.

%

\vspace{5mm}
\facilities{Keck(NIRSPEC)}


\software{Astropy \citep{astropy:2013,astropy:2018}, Batman \citep{batman}, emcee \citep{emcee}, George \citep{george,george2}, iSpec \citep{ispec1,ispec2}, LDTk \citep{LDTK}, Matplotlib \citep{matplotlib}, molecfit \citep{molecfit1,molecfit2}, Numpy \citep{numpy}, p-winds \citep{p-winds_code,DosSantos2022}, REDSPEC \citep{McLean2003,McLean2007}, Scipy \citep{scipy}}



\pagebreak 

\appendix

\section{Detections and non-detections of helium in the exoplanet literature}
\label{sec:lit_detections}

\begin{table*}[ht!]
\centering
\caption{The published detections and robust upper limits of exoplanetary helium absorption in the literature as plotted in Figures \ref{fig:xuv_vs_delta_H} and \ref{fig:ND}. The first reference in the reference column is the reference from which the values are taken or derived. Additional references to studies of these planets are also given.}
\label{tab:lit_he}
\begin{tabular}{c|c|c|c} \hline
Planet & $\mathrm{F_{XUV}}$ (W\,m$^{-2}$) & $\delta R_P/H_{\mathrm{eq}}$ & References \\ \hline
WASP-69b & $4.170 \pm 0.566^a$ & $85.5 \pm 3.6$ & \cite{Nortmann2018}, \citep[also][]{Vissapragada2020} \\
HD\,189733b & $16.75 \pm 0.028^a$ & $77.2 \pm 4.8$ & \cite{Nortmann2018}, \citep[also][]{Salz2018,Guilluy2020} \\
HD\,209458b & $1.004 \pm 0.284^a$ & $46.9 \pm 4.8$ & \cite{Alonso-Floriano2019}, \citep[also][]{Nortmann2018} \\
HAT-P-11b & $2.109 \pm 0.124^a$ & $103.4 \pm 11.3$ & \cite{Allart2018}, \citep[also][]{Mansfield2018} \\
WASP-107b & $2.664 \pm 1.05^a$ & $88.7 \pm 2.1$ & \cite{Kirk2020}, \citep[also][]{Spake2018,Spake2021,Allart2019} \\
GJ\,436b & $0.197 \pm 0.007^a$ & $ \leq 37.5$ & \cite{Nortmann2018} \\
KELT-9b & $\leq 0.15^a$ & $ \leq 41$ & \cite{Nortmann2018} \\
WASP-127b & $0.058 \pm 0.034^a$ & $ \leq 18.77$ & \cite{dosSantos2020} \\
GJ\,1214b & $0.3^{+0.6, a}_{-0.2}$ & $57 \pm 10$ & \cite{Orell-Miquel2022},  \\ & & & \citep[also][]{Kasper2020,Petit2020}\\
GJ\,9827d & $2.4^{+4.8,a}_{-1.6}$ & $\leq 17$ & \cite{Kasper2020}, \citep[also][]{Carleo2021} \\
HD\,97658b & $1.1^{+2.2,a}_{-0.73}$ & $\leq 74$ & \cite{Kasper2020} \\
GJ\,3470b & $1.435 \pm 0.008^a$ & $77 \pm 9$ & \cite{Palle2020}, \citep[also][]{Ninan2019} \\
55\,Cnc\,e & $7.4^{+14.8,b}_{-4.9}$ & $\leq 11$ & \cite{Zhang2020} \\
HAT-P-18b & $8^{+16,b}_{-5}$ & $14.3 \pm 3.5$ & \cite{Paragas2021} \\
HD\,73583b/TOI-560b & $5.1 \pm 1.3^b$ & $123 \pm 10$ & \cite{Zhang2022_TOI} \\
HAT-P-32b & $90 \pm 12^a$ & $72.6 \pm 3.4^c$ & \cite{Czesla2022} \\
WASP-80b & $6.281^{+6.281,b}_{-3.141}$ & $\leq 39$ & \cite{Fossati2021} \\
WASP-76b & $\leq 94^a$ & $\leq 35$ & \cite{Casasayas-Barris2021} \\
GJ\,9827b & $37^{+74,b}_{-25}$ & $\leq 83$ & \cite{Carleo2021} \\
TRAPPIST-1b & $3 \pm +0.4^{b,d}$ & $\leq 1.6$ & \cite{Krishnamurthy2021} \\
TRAPPIST-1e & $0.4 \pm 0.07^{b,d}$ & $\leq 4.2$ & \cite{Krishnamurthy2021}\\
TRAPPIST-1f & $0.27 \pm +0.04^{b,d}$ & $\leq 1.5$ & \cite{Krishnamurthy2021}\\
\hline
WASP-52b & $24.8^{+49.7,a}_{-16.6}$ & $66 \pm 5$ & This work \\
WASP-177b & $3.5^{+7.0,a}_{-2.3}$ & $\leq 22$ & This work \\
\hline
\multicolumn{4}{l}{$^a$ for $\lambda < 504$\,\AA. $^b$ for $\lambda < 912$\,\AA. $^c$ error calculated assuming same fractional uncertainty as in the equivalent width.}\\
\multicolumn{4}{l}{$^d$ calculated from \cite{Wheatley2017}.}\\
\end{tabular}
\end{table*}

\section{The sigma-clipping of frames}
\label{sec:frame_clipping}

As described in Section \ref{sec:obs}, we opted to exclude certain outlying frames from our analyses due to a combination of poor observing conditions and cosmic rays.

\begin{figure}
    \centering
    \includegraphics[scale=0.6]{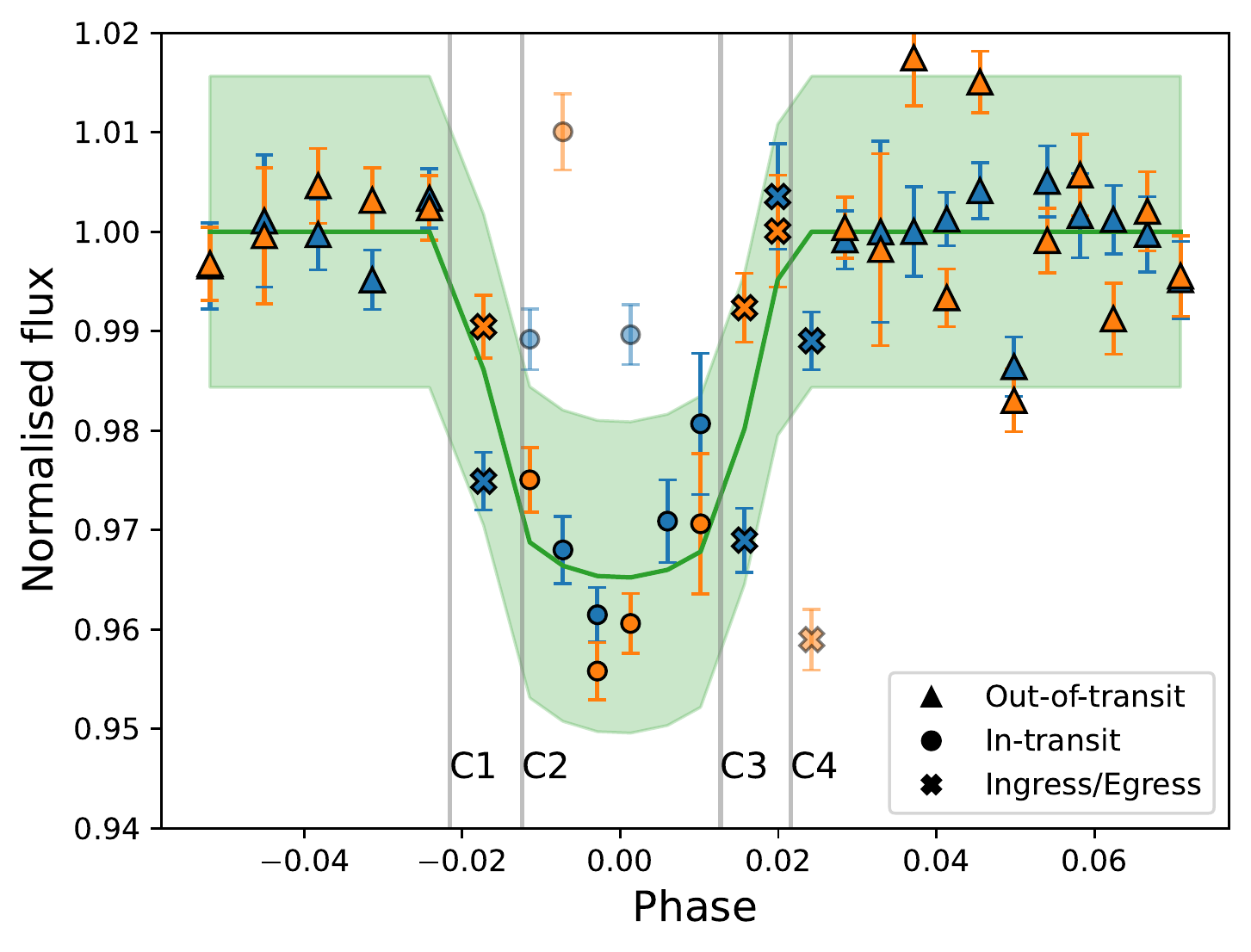}
    \caption{The frame clipping for WASP-52b. The transit light curves are shown for order 70 (blue) and order 71 (orange), calculated in a 0.43\,\AA-wide bin centered on the redder two lines of the helium triplet. These light curves are given in terms of the \textit{excess} transit absorption. The green line shows a fit to the combined data. The shaded green region indicates $\pm3$ median absolute deviations from the model. Any frame falling outside this region was removed (and is shown at a lower opacity). The plot symbols show how we defined the various stages of the transit, with the first-to-fourth contact points labeled. We note that frame 11 for order 71 fell off the bottom of this figure. This figure also demonstrates the repeatibility of our signal in the two separate orders.}
    \label{fig:frame_clipping_w52}
\end{figure}

\begin{figure}
    \centering
    \includegraphics[scale=0.6]{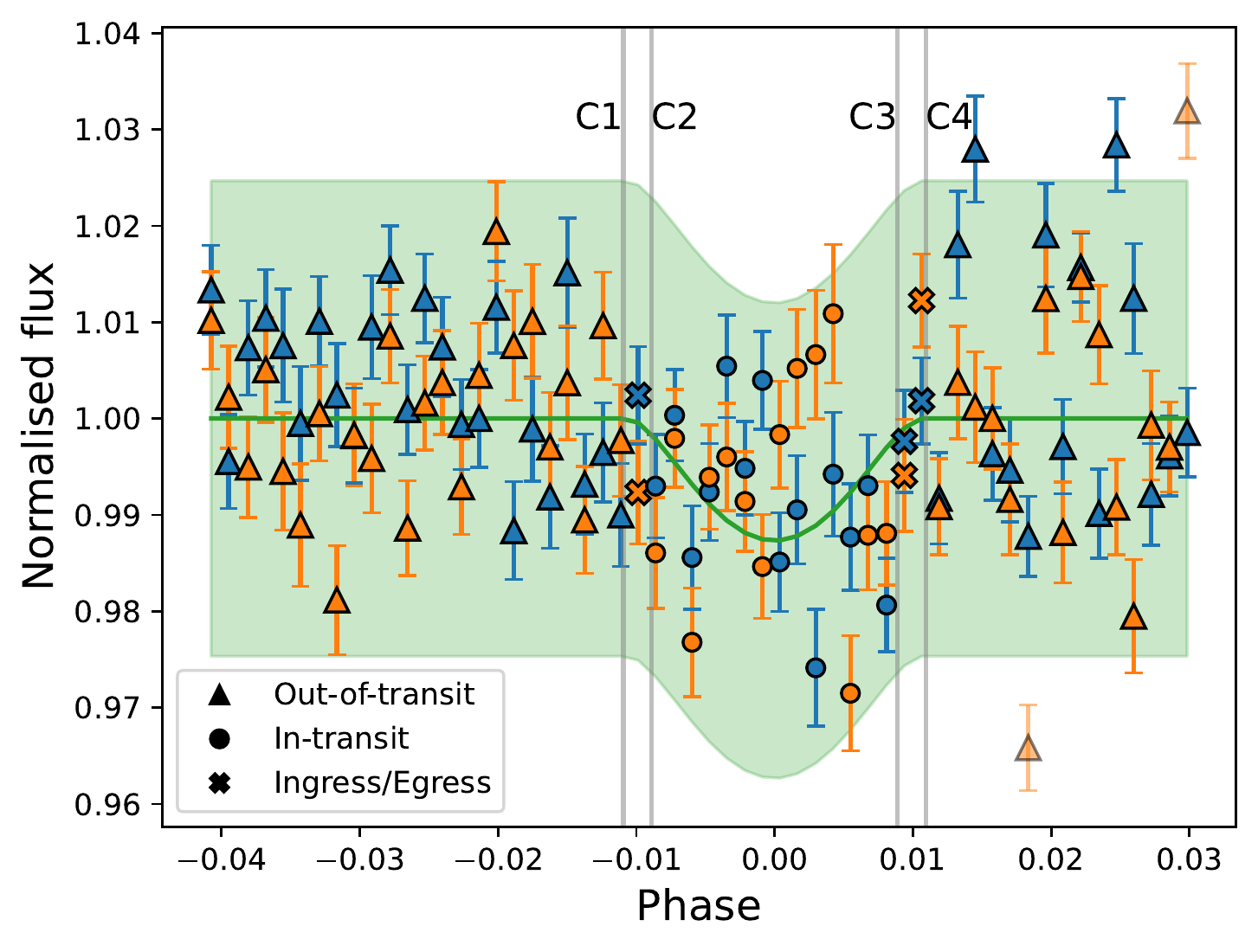}
    \caption{The frame clipping for WASP-177b. See Figure \ref{fig:frame_clipping_w52} for details.}
    \label{fig:frame_clipping_w177}
\end{figure}

We created transit light curves with our data in 0.43\,\AA-wide bins (equal to one resolution element) centered on the mean of the redder two lines of the \ion{He}{1} triplet (10833.261\,\AA) for orders 70 and 71 separately. We then fitted an analytic transit light curve following the procedure described in Section \ref{sec:lightcurves} to the resulting light curves. We excluded those frames that lay $> 4$ median absolute deviations away from this fitted model. Figure \ref{fig:frame_clipping_w52} shows this fitted model along with the frames that were rejected. By this method we rejected frames 7 and 10 from order 70, and frames 8, 11, and 15 from order 71 (10\,\% of the total frames) for WASP-52b. For WASP-177b, we excluded no frames from order 70 and two frames from order 71 (47 and 56, 2\,\% of our spectra), as shown in Figure \ref{fig:frame_clipping_w177}. Figures \ref{fig:frame_clipping_w52} and \ref{fig:frame_clipping_w177} also demonstrate which frames were used to define the in-transit, out-of-transit, ingress, and egress frames.

\bibliography{He_bib}{}
\bibliographystyle{aasjournal}


\end{document}